\documentclass[fleqn]{2023SCGE}
\usepackage[dvipsnames]{xcolor}
\usepackage{orcidlink} 
\usepackage{xspace}
\usepackage{aas_macros}

\def\mgii{Mg\,{\sc ii}\xspace}
\def\mgi{Mg\,{\sc i}\xspace}
\def\hi{H\,{\sc i}\xspace}

\def\hsla{\rm HSLA\xspace}
\def\hifpi{\rm HI4PI\xspace}
\def\siiv{Si\,{\sc iv}\xspace}
\def\siii{Si\,{\sc ii}\xspace}
\def\siiii{Si\,{\sc iii}\xspace}
\def\civ{C\,{\sc iv}\xspace}
\def\ovi{O\,{\sc vi}\xspace}
\def\aliii{Al\,{\sc iii}\xspace}
\def\caii{Ca\,{\sc ii}\xspace}

\def\nai{Na\,{\sc i}\xspace}
\def\feii{Fe\,{\sc ii}\xspace}
\def\nv{N\,{\sc v}\xspace}

\def\kms{km~s$^{-1}$}

\begin{document}
\ensubject{subject}

\ArticleType{Article}
\SpecialTopic{SPECIAL TOPIC: }
\Year{2025}
\Month{June}
\Vol{X}
\No{1}
\DOI{10.1007/s11433-025-2965-5}
\ArtNo{000000}
\ReceiveDate{June 14, 2025}
\AcceptDate{April 7, 2026}
\OnlineDate{May 25, 2026}
\renewcommand\floatpagefraction{.9}
\renewcommand\topfraction{.9}
\renewcommand\bottomfraction{.9}
\renewcommand\textfraction{.1}
\setcounter{totalnumber}{50}
\setcounter{topnumber}{50}
\setcounter{bottomnumber}{50}

\title{The Vertical Structure and Asymmetry of \mgii-enriched Gas in the Milky Way Disk}

\author[1]{Xiaochuan Jiang\orcidlink{0000-0002-3878-5590}}{}
\author[2]{Taotao Fang\orcidlink{0000-0002-2853-3808}}{{fangt@xmu.edu.cn}}
\author[2]{Shulan Yan\orcidlink{0000-0002-3940-2950}}{}

\address[1]{School of Information Engineering, Fujian Business University, Fuzhou, Fujian 350506, China}
\address[2]{Department of Astronomy, Xiamen University, Xiamen, Fujian 361005, China}

\AuthorMark{Jiang Xiaochuan}

\AuthorCitation{Jiang X., Fang T., and Yan S.}

\abstract{
The physical properties of Milky Way \mgii-bearing gas remain poorly constrained due to the saturation of the near-UV doublet. We utilize the weaker \mgii $\lambda\lambda$1239, 1240 doublet from 482 archival HST/COS extragalactic sightlines to probe this cool gas phase. We identify 43 low-velocity absorbers ($|v_{\rm LSR}|<40\ {\rm km\ s^{-1}}$), yielding a covering fraction ($C_f$) of $32\pm5\%$ for $\log N_{\rm MgII} > 15$. We find that $C_f$ follows an exponential decay relative to equivalent width thresholds, marking a transition from a diffuse medium to localized, dense structures (e.g., cold neutral medium cores). The steep decline of the distribution at high column densities likely reflects the saturation of the turbulent log-normal spectrum and dust depletion. By integrating stellar data, we derive a \mgii scale height $h_{\rm MgII} = 0.12\pm0.02\ \rm\ kpc$ and mid-plane density $n_{0,\rm MgII} = (3.9\pm0.4)\times 10^{-6}\ \rm cm^{-3}$. A pronounced north-south asymmetry exists, with the northern hemisphere displaying a significantly higher mid-plane density ($n_{0,n} \approx 4.7 \times 10^{-6}\ \rm cm^{-3}$) than the south ($3.2 \times 10^{-6}\ \rm cm^{-3}$). This discrepancy suggests that the northern interstellar medium is more spatially concentrated and clumpy, whereas the southern gas is more ubiquitously distributed with a lower average density. These results indicate that \mgii is tightly confined to the disk, governed by a unified depletion law and restricted vertical extent.
}



\maketitle

\begin{multicols}{2}

\section{Introduction}\label{sec:intro}

Magnesium (Mg) is a relatively abundant element in the universe, designating a solar abundance ratio of Mg/H = 10$^{-4.4}$ \cite{asplund2009chemical}. 
99\% of Mg originates from exploding massive stars, and the remaining 1\% from exploding white dwarfs \cite{johnson2019populating}. These explosive events generate powerful galactic outflows that expel Mg and other metals from stellar interiors into the interstellar and circumgalactic medium, enriching the gaseous environments surrounding galaxies. 

The dominant form of gas-phase Mg is \mgii in cool
\Authorfootnote

\noindent neutral gas \cite{fitzpatrick1997abundance,jensen2007variation,gnat2007time}.  The ionization energy of \mgi (7.65 eV) lies below the ionization edge of hydrogen (13.6 eV), and that of \mgii (15.04 eV) lies above the Lyman edge. The gas traced by \mgii is cooler than the warm gas ($10^5<T<10^6\ \rm K$) traced by \siiv, \civ, and \ovi \cite{savage2009extension,wakker2012Characterizing}.
Therefore, the \mgii lines are well suited to tracing the cool gas with densities of $n_{\rm H}\approx10^{-2}-10^{-1} {\rm\ cm^{-3}}$ and temperatures of $T\approx10^{3.8}-10^{4}$ K (e.g., \cite{bergeron1994hubble, wolfire1995neutral, ferriere2001interstellar, chen2017mg}). 

Observations of external galaxies reveal that cool gas traced by the near-ultraviolet (NUV) \mgii $\lambda\lambda2796, 2803$ absorption lines extends to $\sim200\ \rm\ kpc$ in projection from galaxies \cite{kacprzak2008halo,chen2010empirical,nielsen2013magiicat}.
Equivalent widths of \mgii and other ions such as \hi, \civ, \siiv, and \ovi show an anti-correlation with impact parameter \cite{Lanzetta1990intermediate,bergeron1991sample,steidel1995nature,bouche2006new,kacprzak2008halo,chen2010empirical}.
Lan 2020~\cite{lan2020coevolution} suggested that the covering fraction $C_f$ of strong \mgii absorbers ($\mathcal{W}_{2796}>1$ \AA) in galaxies evolves with redshift, similar to that of the star formation rate (SFR). 
In particular, the $C_f$ of \mgii absorbers within 0.3 virial radii is significantly higher around star-forming galaxies than around passive galaxies at all redshifts.
These findings indicate a strong connection between \mgii enrichment in galactic halos and star formation in galaxy disks \cite{hopkins2006evolution}. 

However, obtaining accurate column densities for \mgii NUV absorbers in current extragalactic studies is often limited by line saturation and insufficient instrumental resolution (e.g., in large surveys like SDSS \cite{york2000sdss} or DESI \cite{desi2016desi}). Consequently, in many cases only equivalent widths are accessible, limiting kinematic constraints.
The Milky Way provides an ideal environment to study the spatial distribution and physical properties of \mgii absorbers, thanks to a large sample of extragalactic sightlines observed with HST/COS at high resolution ($\rm R\sim15{,}000$).

Herenz et al. (2013) \cite{herenz2013milky} observed the NUV \mgii doublets $\lambda\lambda2796, 2803$ in the Milky Way over a wide velocity range using the {\it HST}/Space Telescope Imaging Spectrograph (STIS). They derived a covering fraction of $\sim0.3$ for high-velocity clouds (HVCs, $|v_{\rm LSR}| > 90$ km s$^{-1}$) with equivalent width $\mathcal{W}_{2796}>0.3\rm\ \AA$. While the total covering fraction of \mgii is near unity due to the ubiquitous low-velocity gas, these strong NUV transitions are often saturated and thus ineffective for accurately measuring gas column density.

The rest wavelengths of the FUV \mgii lines are $\lambda=1239.93$ and $1240.39$ \AA. Their oscillator strengths are $6.32\times10^{-4}$ and $3.56\times10^{-4}$, which are thousands of times lower than those of the NUV doublet \cite{morton2003atomic}. 
Available in the far ultraviolet (FUV), the weaker \mgii transitions $\lambda\lambda1239,\ 1240$, because their oscillator strengths are lower and remain unsaturated, can better constrain the column density of \mgii-enriched gas \cite{morton2003atomic}. The \mgii gas phase distribution is also influenced by \mgii depletion. 
Jensen \& Snow 2007 \cite{jensen2007variation} studied \mgii $\lambda\lambda1239, \ 1240$ absorption along 44 OB stellar sightlines at Galactic latitudes of $|b|<20$ using {\it HST}/STIS and showed that the \mgii depletion pattern correlates with hydrogen density, molecular hydrogen, and extinction along the lines of sight.
Mg depletion correlates with hydrogen density in HST/Goddard High-Resolution Spectrograph studies \cite{savage1996interstellargas, destree2010detection, fitzpatrick1997abundance, cardelli1995gas}.

In this work, we use a large sample of archival {\it HST}/COS sightlines to study the weak \mgii doublet absorption 1239/1240 \AA\ in the FUV. By accurately measuring the column densities of these \mgii absorbers, we investigate the spatial distribution and physical properties of \mgii-enriched gas around the Milky Way.
The manuscript is organized as follows: Section 2 covers our data reduction methods and analysis. Sections 3 and 4 present the \mgii distribution and discussion, respectively, followed by a summary in Section 5.

\section{Data \& Analysis}

We obtained FUV spectra from the Hubble Spectroscopic Legacy Archive (\hsla, \cite{peeples2017hubble}). As of September 2023, there were 2008 sources in the \hsla, including planets, stars, galaxies, and clusters. The \hsla provides co-added {\it HST}/COS spectrum with G130M, G140L, and G160M grating, covering Ly$\alpha$ and FUV \mgii lines. 

We identified a sample of 482 extragalactic sightlines with a spectral signal-to-noise ratio (SNR) greater than 6 per resolution element. The SNR was calculated over a 10 \AA\ absorption-line-free window from 1235 to 1245 \AA, covering the \mgii doublet lines. The SNR for each sightline is tabulated in Table~\ref{lqso}.
These extragalactic sightlines were typically proposed to study the properties of halo gas in extragalactic systems or high-velocity clouds in the Milky Way; hence, the sample is not preferentially selected for low-velocity absorbers. 

\subsection{{\it HST}/COS Data Reduction}\label{fit}
For each {\it HST}/COS spectrum downloaded from \hsla, we examined a 10 \AA\ spectral window centered around 1239 \AA\ to search for potential \mgii absorbers within velocities of $|V_{LSR}|<500\rm\ km\ s^{-1}$, where LSR stands for the Local Standard of Rest.
This interval, ranging from 1235 \AA\ to 1245 \AA, includes a strong \nv doublet at 1238.82 \AA\ and 1242.80 \AA, near the \mgii lines. The wavelength offset between the \nv doublet lines is much larger than that of the \mgii lines, allowing \nv to be distinguished from \mgii within this interval.

We identified the Galactic \mgii $\lambda\lambda1239, \ 1240$ doublet based on their characteristic wavelength separation. To minimize contamination from intergalactic medium (IGM) lines (e.g., random Ly$\alpha$ forest lines), we required that both components of the doublet be detected at the same velocity (within the typical Galactic range) and exhibit consistent kinematic profiles. Systems showing significant discrepancies in line profiles or unphysical doublet ratios, indicative of IGM blending, were excluded.

We binned each spectrum by 3 pixels to boost the SNR while retaining adequate sampling. For each detected \mgii absorption, we fit it with a Voigt profile (VP). First, we masked significant absorption features, including \mgii absorption, within the 10 \AA\ spectral window, and fit the continuum using the \texttt{UnivariateSpline} function from \texttt{Scipy}. To account for the uncertainty in the continuum placement, we estimated the noise level using the root-mean-square (RMS) of the flux in the line-free regions. We then fit the \mgii doublet absorption at 1239 and 1240 \AA\ simultaneously with a VP convolved with the COS G130M line-spread function \cite{ghavamian2009cos} using \texttt{Lmfit}\footnote{https://github.com/lmfit/lmfit-py}. This fitting yields column density $N$, Doppler parameter $b$, and heliocentric velocity $v$, which we convert to the LSR frame adopting the Standard Solar Motion parameters $(U, V, W) = (9, 12, 7)$ km s$^{-1}$ defined in \cite{mihalas1981galactic}: 
\begin{equation}
v_{\rm LSR}=v_{hel} + 9 \cos(l)\cos(b)+12 \sin(l)\cos(b) + 7\sin(b) \nonumber
\end{equation}

\begin{figure}[H]
\centering
\includegraphics[width=1.0\columnwidth]{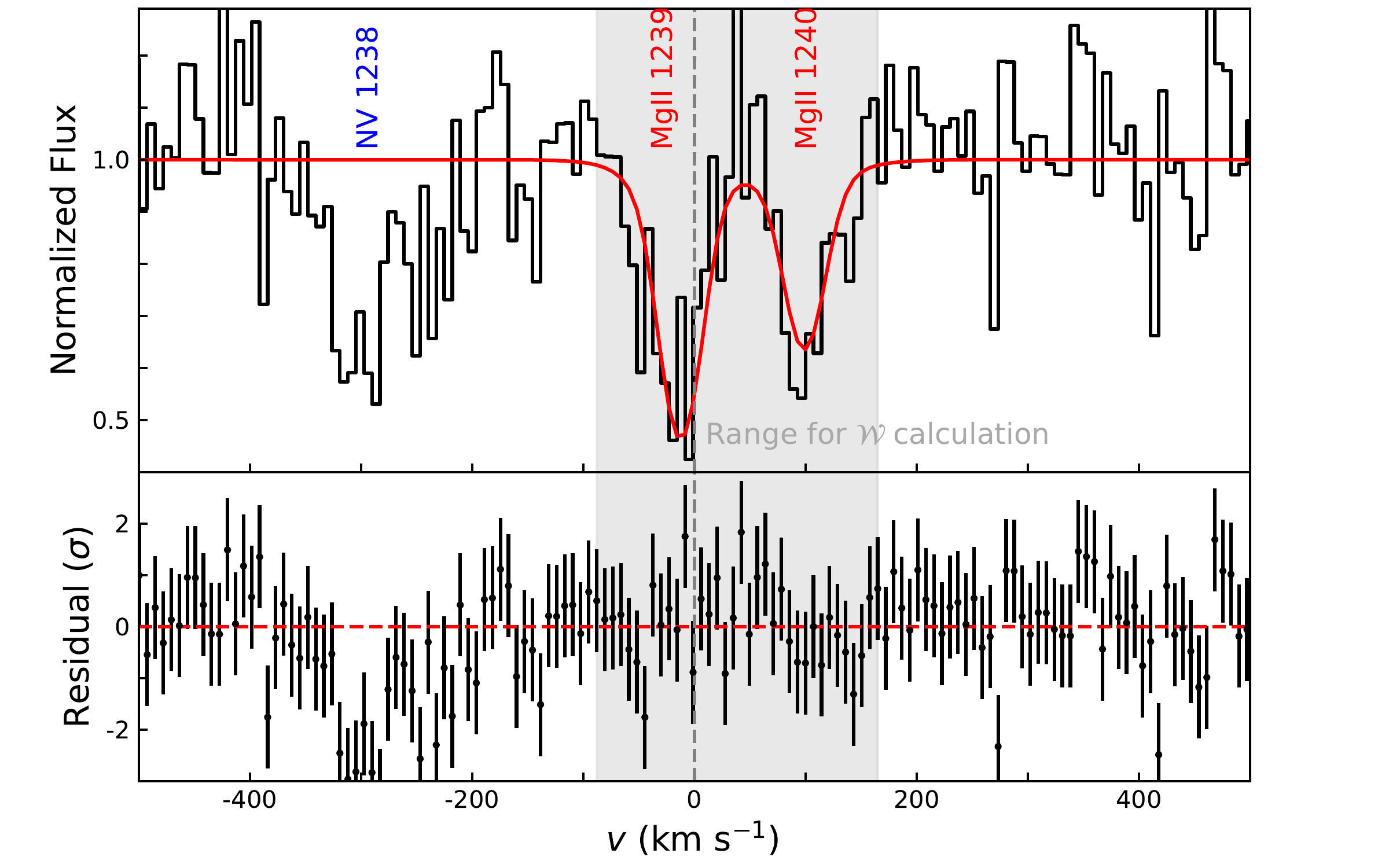}
\caption{The normalized spectrum of PDS 456 in the velocity space of \mgii $\lambda1239$ and the residuals. The \mgii column density along this source is the largest in our sample. The red line in the upper panel represents the best-fit VP. The wavelength range of equivalent width calculation is shown in the gray region.} 
\label{fitres}
\end{figure}

Figure~\ref{fitres} presents an example of the normalized spectra and the convolved VP of the \mgii absorber in the spectrum of PDS 456.
We calculated the equivalent widths of the \mgii lines directly from the continuum-normalized spectra.
The wavelength range for this calculation is determined by the VP fit: we include pixels where the modeled absorption exceeds 1\% (e.g., the gray region in Figure~\ref{fitres}).

Figure~\ref{fitres} represents the strongest absorber in our sample (towards PDS 456), where the wings of the doublet members partially blend. This blending is primarily driven by the extended wings of the instrumental line spread function (LSF) and Gaussian broadening, rather than true damping wings. The equivalent width ratio of the two lines for this specific absorber is approximately 1.8:1, implying only modest unresolved saturation. While it would be feasible to deblend the lines for this exceptionally strong absorber, this sightline is not representative of the rest of the dataset, which consists of much weaker absorbers. Therefore, to maximize the signal-to-noise ratio for the weak absorbers and to ensure a consistent methodology across the entire sample, we measured the combined equivalent width of the doublet ($W_{1239} + W_{1240}$) for all targets.

Subsequently, to determine the \mgii column density using the COG method, we employed a theoretical curve of growth constructed for the total equivalent width of the doublet. We adopted a typical Doppler parameter of $b=10\rm\ km\ s^{-1}$ \cite{Roamn-Duval2021Metal}, which is generally consistent with our VP fitting results. The spectral resolution of {\it HST}/COS using grating G130M varies with the detector lifetime position, resulting in an LSF with a FWHM of approximately $15-20\rm\ km\ s^{-1}$ \cite{soderblom2021cos}. This corresponds to an instrumental Doppler broadening parameter of $b_{\rm inst} \approx 9 - 12\rm\ km\ s^{-1}$. Consequently, absorption lines with $b \lesssim 10\rm\ km\ s^{-1}$ are dominated by the instrumental LSF and remain essentially unresolved.

Because VP fitting inherently yields unconstrained $b$-value errors for these unresolved weak lines due to strong $b-N$ degeneracy, adopting a fixed representative $b$-value is a stable approach. To robustly account for the uncertainty in this assumption, we treated the natural variation of the Doppler parameter as a systematic uncertainty. Based on the typical $1\sigma$ dispersion of resolved \mgii components in high-resolution studies (e.g., \cite{churchill2020mgii}), we assigned an empirical uncertainty of $\Delta b = \pm 4\rm\ km\ s^{-1}$ to our assumed $b=10\rm\ km\ s^{-1}$. We then performed a Monte Carlo simulation ($1000$ iterations) to propagate both the statistical EW measurement errors and the systematic $b$-value uncertainty into the final column density errors. During the sampling, we imposed a physical lower limit of $b_{\rm min} = 2.6\rm\ km\ s^{-1}$ (corresponding to the pure thermal broadening of \mgii at $T = 10^4\rm\ K$) and an upper limit of $b_{\rm max} = 25\rm\ km\ s^{-1}$ to exclude unphysical single-component kinematics. The resulting combined uncertainties, which properly reflect the impact of modest saturation, are reported in Table~\ref{lqso}.

\begin{figure}[H]
    \centering
    \includegraphics[width=0.9\columnwidth]{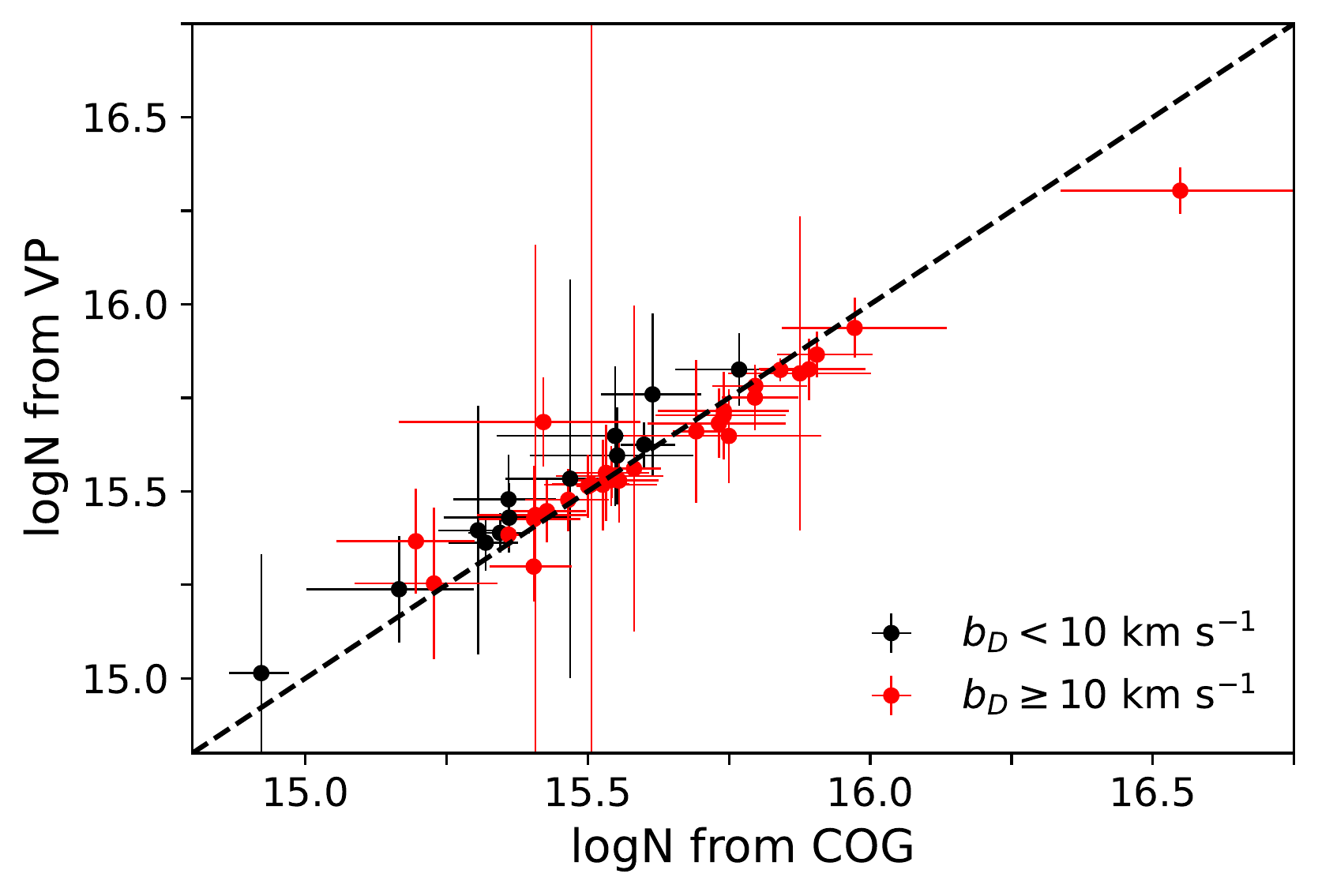}
    \caption{The figure presents a comparison of \mgii column densities obtained using the COG and VP methods. The red and black points denote absorbers with $b\geq10\rm\ km\ s^{-1}$ and $b<10\rm\ km\ s^{-1}$, respectively. The diagonal line indicates the equivalence of column densities for the two quantities.}
    \label{fig:cogvp1}
\end{figure}

We compare \mgii column densities using the COG and VP methods, as shown in Figure~\ref{fig:cogvp1}. Overall, the results are consistent. The scatter is driven by the deviation of the fitted $b$ values from the fixed $b=10\rm\ km\ s^{-1}$ used in the COG. Specifically, absorbers with fitted $b < 10\rm\ km\ s^{-1}$ tend to have $N_{\rm VP} > N_{\rm COG}$, while those with $b > 10\rm\ km\ s^{-1}$ show the opposite trend. This suggests that $b=10\rm\ km\ s^{-1}$ is a representative characteristic value for our sample.

For the majority of absorbers ($\log N < 16$), the lines are weak and lie on the quasi-linear part of the curve of growth. Since most of these lines are unresolved ($b \lesssim b_{\rm inst}$), the VP method is subject to degeneracy between $b$ and $N$. Therefore, we adopt the COG method for absorbers with $N < 10^{16}\, \rm\ cm^{-2}$ to ensure robustness.

Only one absorber in our sample (towards PDS 456) exhibits a column density exceeding $10^{16}\, \rm\ cm^{-2}$. For this saturated line, the COG method with a fixed $b$ significantly overestimates the column density. Thus, we use the VP method for this single strong absorber to properly account for the saturation wings.

\begin{figure}[H]
    \centering
    \includegraphics[width=0.9\columnwidth]{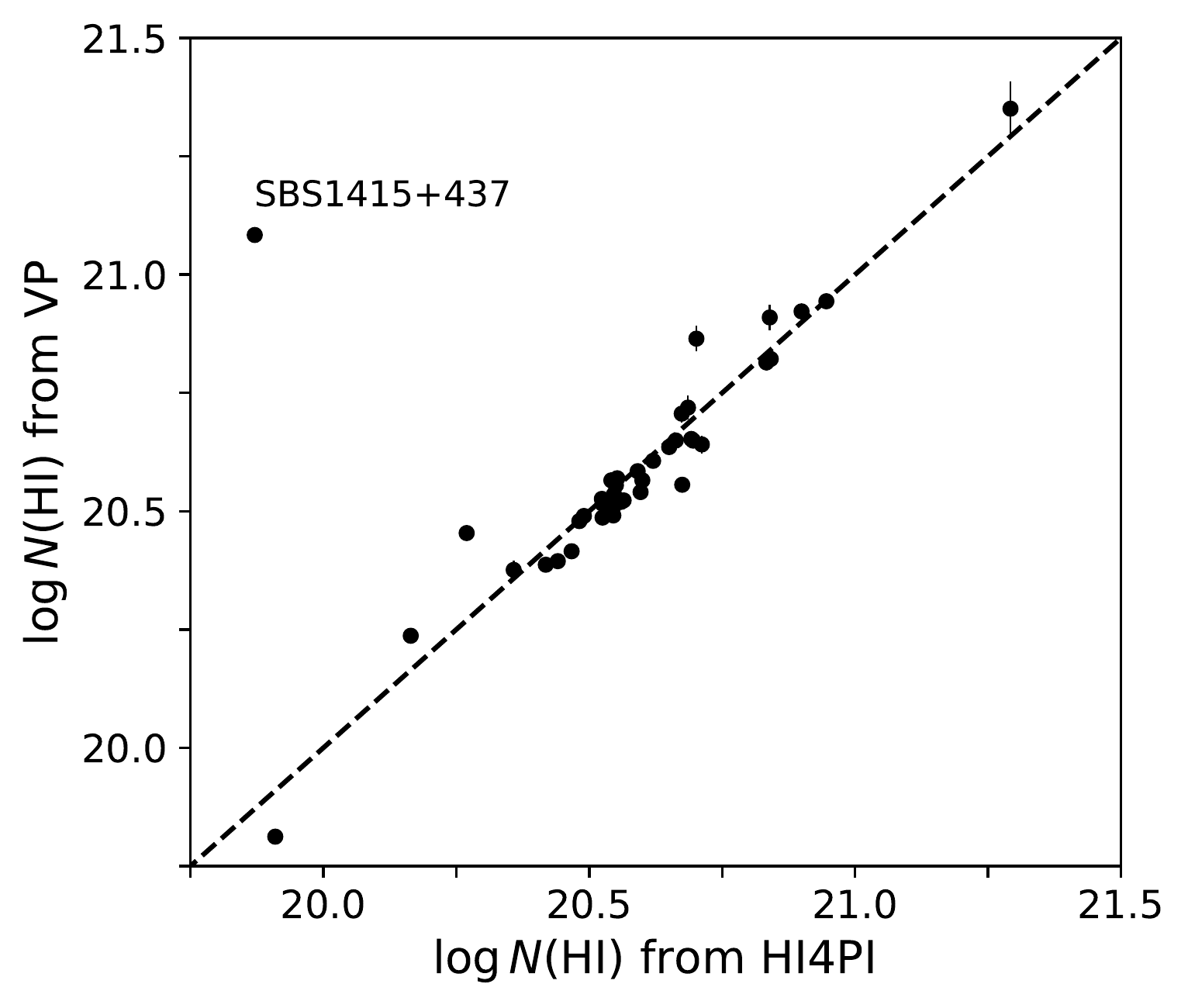}
    \caption{The figure displays the comparison of \hi\ column densities obtained via \hifpi\ and VP methods. The diagonal line indicates the equivalence of column densities for the two quantities.}
    \label{fig:cogvp2}
\end{figure}

In our analysis, we only consider \mgii absorption lines with a line significance exceeding $3\sigma$. The COS spectrograph, using the G130M grating, captures regions containing significant \siii and \siiii lines. The abundance of silicon is similar to that of magnesium, and \siii and \siiii are the predominant forms of silicon at temperatures $T \lesssim 10^{4}$ \cite{gnat2007time}. 
The ionization potential of \siii (16.4 eV) is quite similar to that of \mgii (15.0 eV), indicating that both ions may trace the same gas phase.

Moreover, the intrinsic strength of the \siii and \siiii lines is greater than that of the NUV \mgii. 
Since $\sim$70\% of silicon and 99\% of magnesium originate from exploding massive stars (with the remainder from exploding white dwarfs; \cite{johnson2019populating}), their shared nucleosynthetic origin implies both elements should co-exist in enriched gas.
Consequently, regions with detectable \mgii absorption should inherently contain silicon.

Given that \siii and \siiii absorption lines are typically stronger and often exhibit multi-component structures or saturation, we did not impose a strict peak-to-peak velocity difference threshold. Instead, we required that the velocity centroid of the \mgii candidate falls within the velocity range spanned by the significant absorption of the corresponding \siii or \siiii lines.
We therefore rule out four \mgii candidates lacking corresponding \siii/\siiii lines.

Finally, we identified 43 low-velocity \mgii absorbers from 482 extragalactic sightlines and present the physical parameters of the \mgii lines in Table~\ref{lqso}.
The median $v_{\rm LSR}$ and $b$ of the sample are $-7.28\pm0.87\ \rm km\ s^{-1}$ and $13.45\pm1.57\ \rm km\ s^{-1}$, respectively. The medians are estimated using the bootstrap method. All errors in this paper correspond to the 68\% confidence interval.

\begin{figure*}
\centering
\includegraphics[width=0.6\textwidth]{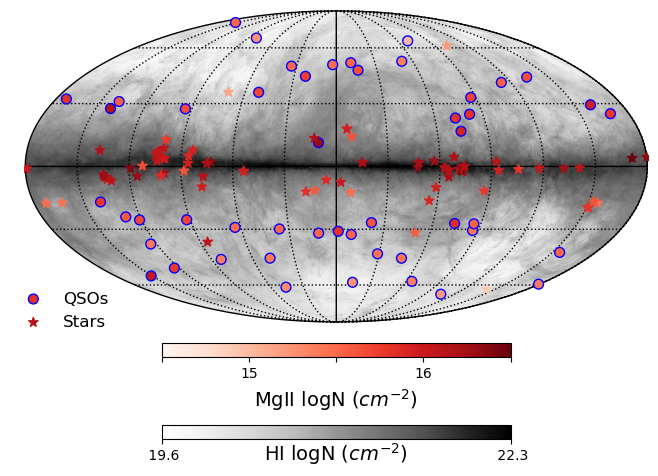}
\caption{The \mgii sightlines are plotted on the all-sky column density map of \hi\ gas in Galactic coordinates using the Mollweide projection. The column density of \hi\ gas is integrated over the velocity range from $-90\leq v_{\rm LSR} \leq 90\ \rm km\ s^{-1}$ \cite{lenz2017new}. The colors of the background and symbols indicate the \hi\ and \mgii column densities, respectively.} 
\label{fig:sky}
\end{figure*}

\subsection{{\rm \hi} Data Reduction}
The depletion of \mgii significantly influences its distribution in the gas phase. Previous studies suggest that the hydrogen column density, which includes both \hi\ and H$_2$, is closely associated with the depletion process (e.g. \cite{Murray1984Interstellar,Jenkins1986Abundances,jensen2007variation}).

To obtain the \hi\ column density, we analyze the COS spectrum covering the Ly$\alpha$ absorption lines. We implement a third-degree polynomial fit for the continuum spectrum between 1195 \AA\ and 1235 \AA, employing VP analysis for the \hi\ absorption lines. 
During this process, we masked the spectral pixels contaminated by geocoronal Ly$\alpha$ emission and constrained the \hi\ column densities by fitting the extended damping wings.
Subsequently, we conduct a simultaneous fit of the continuum and absorption lines to determine the \hi\ column density, as listed in Table~\ref{hi}. These VP-derived column densities will be utilized in Section 4.1 to calculate the total hydrogen column density.

The table also presents the Galactic coordinates of the sightlines, along with the \hi column density from the \hi\ 4-PI Survey (\hifpi, \cite{bekhti2016hi4pi}) and the H$_2$ estimates from Kalberla et al. (2020) \cite{Kalberla2020HI}.
It is important to note the difference in spatial sampling: the \hifpi survey measures emission averaged over a $16.2^\prime$ beam, whereas the HST/COS absorption spectroscopy probes a `pencil beam' line of sight toward the point-like background AGNs. The implications of comparing 21-cm emission with Ly$\alpha$ absorption under these differing beam sizes have been discussed in detail by Wakker et al. (2011) \cite{wakker2011measuring}.

We compare the \hi\ column density from the VP method with that from \hifpi in Figure~\ref{fig:cogvp2}. 
The analysis of the Galactic Ly$\alpha$ absorption towards SBS 1415+437 (SDSS J1417+4330) is complicated by the intrinsic spectrum of the source. With a low redshift of $z_{em} \approx 0.002$, the broad intrinsic Ly$\alpha$ emission and associated absorption features of the host galaxy partially overlap with the Galactic profile, affecting the continuum placement and VP fitting.
Excluding the source, the VP-derived column densities align well with those from \hifpi, showing a Spearman correlation of $r_s=0.95\pm0.01$. 
We plotted the global map of \hi 21 cm emissions from \hifpi, highlighting the \mgii sightlines (Figure~\ref{fig:sky}). 

\section{\mgii Distribution}\label{ana}
The identified \mgii absorbers offer valuable insights into the column densities and spatial distribution of the cool gas. In the following discussion, we will explore the position, scale height, covering fraction, and asymmetry of the \mgii gas.

\subsection{Position of {\rm \mgii} gas}

The velocities of all 43 identified absorbers are less than 40 \kms, suggesting a likely interstellar medium (ISM) origin.
Most low-velocity \mgii gas is expected to be close to or in the Galactic disk, similar to other ions like \hi, \aliii, \siiv, and \ovi (e.g., \cite{nielsen2013magiicat,nielsen2013magiicatII,werk2013cos, lehner2015evidence, keeney2017characterizing}). 
This is further supported by the negative correlation between the column density of low-velocity \mgii absorbers and the Galactic latitude $|b|$ (Spearman's correlation $r_s = -0.62\pm0.06$).

However, no high-velocity ($|v_{\rm LSR}| \gtrsim 90\ {\rm km~s^{-1}}$) \mgii absorbers were found. 
Using {\it HST}/STIS spectra, Herenz et al. 2013 \cite{herenz2013milky} identified high-velocity \mgii-bearing gas by \mgii $\lambda\lambda2796,2803$ lines along 11 extragalactic sightlines.
Our analysis includes five of their sightlines; however, we did not observe high-velocity \mgii $\lambda\lambda1239, 1240$ absorption. They reported \mgii column densities of $\log (N/\mathrm{cm}^{-2}) \approx 13.6$ or lower, which are well below our FUV \mgii detection limit for G130M at the available S/N.

\subsection{Covering fraction}\label{sec_comp}

To determine the covering fraction, we calculated the minimum detectable equivalent width ($\mathcal{W}_{lim}$) for each extragalactic sightline using the IDL code COS\_EWLIM\footnote{\url{https://casa.colorado.edu/~danforth/science/cos/cos_ewlim.pro}} at a 3$\sigma$ significance level (see \cite{keeney2012significance} for details). Using $\mathcal{W}_{lim}$, we identified the number of sightlines ($\mathcal{N}_{tot}$) capable of detecting \mgii at a given threshold. From this subset, we counted the sightlines exhibiting \mgii absorption ($\mathcal{N}_{det}$) to derive the covering fraction ($\mathcal{N}_{det}/\mathcal{N}_{tot}$).

Since $C_f$ represents a binomial proportion, we estimated the uncertainties using the Wilson score interval \cite{cameron2011estimation}. Unlike standard Poisson or Gaussian approximations, this method yields asymmetric error bars strictly bounded within the physical range of [0, 1], providing a robust statistical representation for detection rates.
For instance, in our sample of 482 sightlines, 85 provide a detection limit of $\log N_{\rm lim} \le 10^{15}\rm\ cm^{-2}$. Within these 85 sightlines, \mgii absorbers are detected in 27 sightlines, yielding $C_f=32^{+5}_{-5}\%$ for the \mgii absorbers ($\log N>15$).

\begin{figure}[H]
    \centering
    \includegraphics[width=0.9\columnwidth]{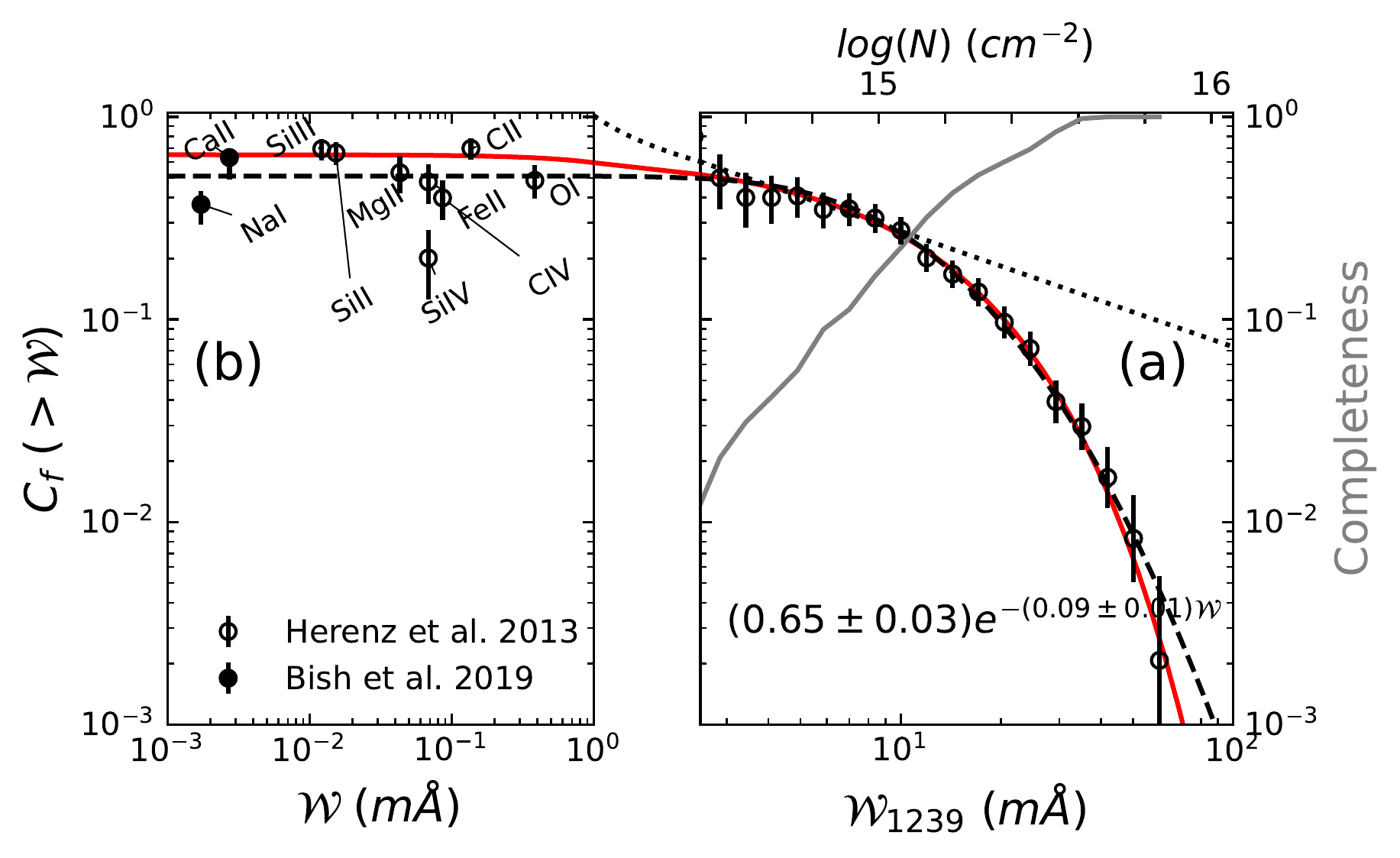}
    \caption{Analysis of the covering fraction ($C_f$). (a) The dependence of $C_f$ on the equivalent width threshold, $\mathcal{W}_{\rm lim}$ (bottom axis) and the corresponding column density $N_{\rm lim}$ (top axis). The black points represent individual measurements with error bars denoting the binomial Wilson score intervals.
The red curve shows the best-fit exponential function ($C_f \propto e^{-N/N_c}$), which provides the statistically preferred description. The black dotted line shows a power-law fit ($C_f \propto N^{1-\beta}$) to the low-threshold data, yielding a slope consistent with cosmic surveys. The black dashed line represents a cumulative log-normal distribution fit ($C_f(>N)$). The gray curve (right axis) indicates the sample completeness, representing the fraction of the 482 sightlines with a detection threshold better than $\mathcal{W}_{\rm lim}$. Note that even at $\mathcal{W}_{\rm lim} \approx 10$ m\AA, the effective sample size ($\approx 100$ sightlines) remains large enough for robust statistical analysis. 
(b) The extrapolated relations of the three models derived from panel (a), plotted against equivalent width. 
For comparison, we overplot the covering fractions of other ions reported by \cite{herenz2013milky} and \cite{bish2019galactic}. Note that the comparison samples from \cite{herenz2013milky} and \cite{bish2019galactic} have different kinematic selection criteria (HVCs and IVCs) compared to this work.}
    \label{ew_fc_tot}
\end{figure}

Figure~\ref{ew_fc_tot}(a) presents the covering fraction as a function of column density threshold, $N_{\rm lim}$. Since the weak FUV transitions generally lie on the linear part of the curve of growth (where $\mathcal{W}_{\rm lim} \propto N_{\rm lim}$), we analyzed the distribution directly in the column density domain.
We compared three functional forms to describe the distribution: a power-law, a cumulative log-normal distribution, and an exponential decay model.

Statistically, the exponential decay model (red curve) provides the best description of the data, yielding the lowest Akaike Information Criterion (AIC$= -20.0$) compared to the cumulative log-normal model (AIC$= -11.6$). The best-fit exponential relation is expressed as:
\begin{equation}
C_f(>N_{\rm lim}) = C \exp\left(-\frac{N_{\rm lim}}{N_c}\right) \label{eq:fw}
\end{equation}
with parameters $C=0.65\pm0.03$ and characteristic column density $\log N_c=15.11\pm0.05$.

Physically, the data reveals a transition in gas properties. Fitting a power-law $C_f(>N_{\rm lim}) \propto N_{\rm lim}^{1-\beta}$ to the low-threshold regime ($N_{\rm lim} < N_c$) yields a shallow differential slope of $\beta = 1.57 \pm 0.08$ (gray dashed line), consistent with the canonical cosmic value for diffuse gas ($\beta \approx 1.6$; e.g., \cite{churchill2003physical}). However, the distribution steepens significantly at higher thresholds ($N_{\rm lim} > N_c$, $\beta = 3.2\pm0.2$), deviating from the cosmic power-law.

We also tested a cumulative log-normal distribution (black dashed line), which typically characterizes the turbulent ISM \cite{vazquez1994hierarchical,ostriker2001density}: 
\begin{equation}
\begin{split}
C_f(>N_{\rm lim}) = \int_{N_{\rm lim}}^{\infty} & \frac{A}{\sqrt{2\pi}\ln(10) \sigma N} \times \\
& \exp\left[ -\frac{(\log_{10} N - \log_{10} N_0)^2}{2\sigma^2} \right] dN
\end{split} \label{eq:lognormal}
\end{equation}
The best-fit parameters are normalization $A = 0.51 \pm 0.05$, characteristic column density $\log_{10} N_0 = 15.08 \pm 0.04$, and width $\sigma = 0.32 \pm 0.02$. While the log-normal model captures the high-column density cutoff, it underestimates the covering fraction at the low-column density end.

Ultimately, the high-column density regime ($\log N \gtrsim 15$) deviates significantly from the canonical power-law extrapolation that typically characterizes diffuse halo gas. This behavior strongly suggests a composite structure for the Galactic \mgii\ gas: whereas the low-column density regime is governed by the pervasive, diffuse halo (power-law), the high-column density end measured in our sample reflects the dense ISM structures in the Galactic disk. This steep cutoff is likely shaped by turbulence and limited by phase transitions or dust depletion in these dense environments.

\begin{figure}[H]
    \centering
    \includegraphics[width=0.9\columnwidth]{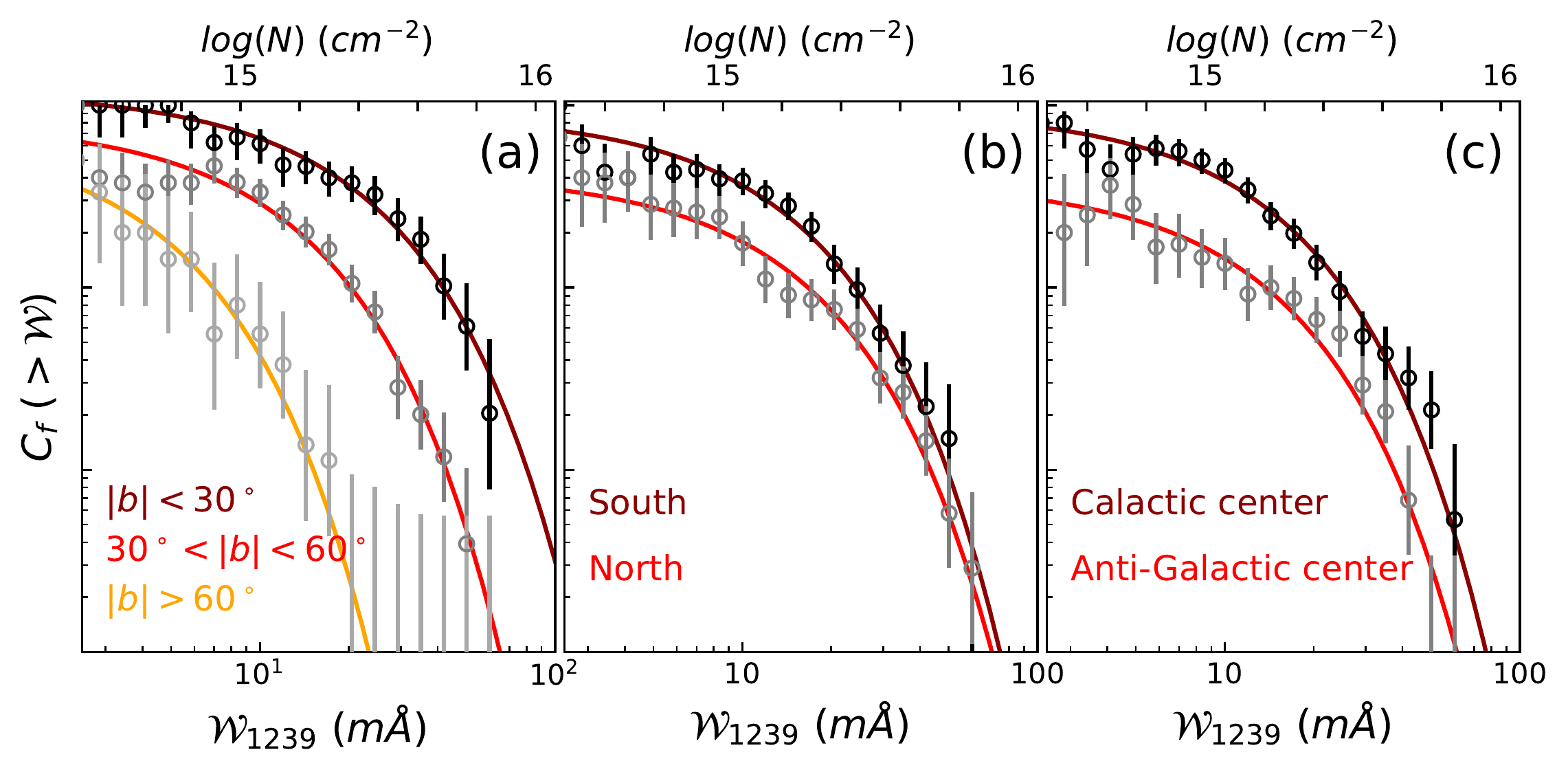}
    \caption{
    Panel (a) presents $C_f$ across three latitude ranges: $|b|\leq 30^\circ$ (black points), $30^\circ\leq |b|\leq 60^\circ$ (gray points), and $|b|\geq 60^\circ$ (light gray points).
    Panel (b) compares $C_f$ between the southern (black points) and northern (gray points) hemispheres, excluding a $\pm45^\circ$ region around the Galactic center in longitude to avoid its influence.
    Panel (c) contrasts $C_f$ in the directions towards the Galactic center (black points; $l < 90^\circ$ or $l > 270^\circ$) and anti-Galactic center (gray points; $90^\circ < l < 270^\circ$).
    In all panels, the curves represent the corresponding exponential fits, with fit parameters annotated at the lower left.
    }
    \label{ew_fc_div}
\end{figure}

The physical distinction between these gas regimes is also reflected spatially. The covering fraction declines significantly as $\mathcal{W}_{\text{lim}}$ increases, and at a fixed $\mathcal{W}_{\text{lim}}$, the covering fraction decreases with increasing latitude from 0-30$^\circ$ to 60-90$^\circ$, as illustrated in Figure~\ref{ew_fc_div}(a).

Our data predominantly probes the dense disk structures, yielding a covering fraction of $C_f \approx 0.32$ at $\log N > 15$. Consequently, comparing this value to literature measurements for other ionized species requires careful consideration of both sensitivity limits and physical origins. For instance, \siii shows $C_f = 0.67$ ($\log N > 12.25$; \cite{herenz2013milky}), while \caii and \nai exhibit $C_f = 0.63$ ($\log N > 11.5$) and $C_f = 0.37$ ($\log N > 11.3$; \cite{bish2019galactic}), respectively. The comparison samples from \cite{herenz2013milky} and \cite{bish2019galactic} explicitly targeted lower-column, diffuse gas selected by kinematics (HVCs and IVCs), which fundamentally differs from the high-column material defining our exponential cutoff.

Nevertheless, we overplot these literature values alongside the extrapolation of Equation~\ref{eq:fw} in Figure~\ref{ew_fc_tot}(b) to illustrate the severe observational bias introduced by detection thresholds. Extrapolating our fit yields an inferred $C_f \sim 0.6$ at $\log N_{\rm lim} > 12.7$, numerically similar to the HVC measures of Herenz et al. (2013). Rather than implying that our dense gas and the literature HVC/IVC gas are physically identical, this extrapolation demonstrates that the seemingly low observed $C_f$ of FUV \mgii\ in our sample is primarily a consequence of our high detection threshold ($\sim10^{15}\ \rm cm^{-2}$) compared to the much higher sensitivity limits of other ion surveys ($\sim10^{11-13}\ \rm cm^{-2}$). It reinforces the composite picture: if observations were sensitive enough to detect the diffuse phase uniformly without kinematic pre-selection, the pervasive halo gas would naturally drive the overall covering fraction to the high values seen in literature surveys.

\subsection{Scale height of \mgii gas}
A simple plane-parallel model describes the column density $N$ as a function of the scale height $z$ and latitude $b$ \cite{savage2009extension}:
\begin{equation}
N = N_\perp(1-e^{-|z|/h})/\sin|b|, \label{equ:mod}
\end{equation}
where $N_\perp=n_0h$, $n_0$ represents the mid-plane density, and $h$ represents the exponential scale height. The parameter $N_\perp$ denotes the column density projected perpendicular to the Galactic plane.

For extragalactic background sources where $z\gg h$, the column density multiplied by the $\sin |b|$ becomes a constant $N_\perp$. We adopt the bootstrapping method with replacement and obtain $\log N_\perp=15.29^{+0.04}_{-0.03}$. 
Additionally, FUV \mgii absorbers have been observed in Milky Way stellar sightlines by STIS \cite{cardelli1995gas,savage1996interstellargas,fitzpatrick1997composition,cartledge2006homogeneity,jensen2007variation,destree2010detection}. From the literature, we compile the total \mgii column density along stellar sightlines, star distances from us, and their heights (see Table~\ref{tab:mg2stars}). Most of the latitudes of these stellar sightlines are below $20^\circ$.

To strictly constrain the disk parameters while accounting for the intrinsic inhomogeneity of the ISM, we adopt a Bayesian approach. We model the data variance as $\sigma_k^2 = e_k^2 + \sigma_p^2$, where $\sigma_p$ is represents the intrinsic scatter, which serves as a logarithmic measure of the spatial patchiness or irregularity in the gas distribution \cite{savage2009extension, wakker2011measuring}.

Crucially, to avoid the bias toward high-density regions caused by the sensitivity-limited nature of absorption surveys, we implement a sensitivity-dependent zero-inflated likelihood. This formulation explicitly models the probability that a non-detection arises either from a physical void (determined by the covering fraction) or from a cloud falling below the detection threshold.

The probabilistic model is constructed in the logarithmic column density space. Let $y = \log_{10}(N)$ be the observed column density and $y_{\rm mod} = \log_{10}(N_{\rm mod})$ be the model prediction, where $N_{\rm mod}$ derived in Equation~\ref{equ:mod}.

For each sightline, we determine the detection probability weight, $C_f$, based on its limiting column density $N_{\rm lim}$ using the empirical relation given in Equation~\ref{eq:fw}. The likelihood function is then formulated to account for both detections and upper limits based on these weights. 

The likelihood for a single sightline is:
\begin{equation}
\mathcal{L} =
\begin{cases}
C_f \cdot \frac{1}{\sqrt{2\pi}\sigma_k} \exp\left[-\frac{(y - y_{\rm mod})^2}{2\sigma^2_k}\right] & \text{for detections}, \\
(1 - C_f) + C_f \cdot \Phi\left(\frac{y_{\rm lim} - y_{\rm mod}}{\sigma_k}\right) & \text{for upper limits},
\end{cases}
\end{equation}
where $\Phi(x)$ is the cumulative distribution function (CDF) of the standard normal distribution, representing the probability that the model-predicted gas column falls below the observed upper limit $y_{\rm lim}$.

Using Markov chain Monte Carlo (MCMC) with \emph{emcee} \cite{Foreman2013emcee}, we determine the posterior probability distributions for the parameters. The resulting best-fit parameters and their 68\% confidence intervals are: $h=0.12\pm0.02\rm\ kpc$, $n_0=(3.9\pm0.4)\times10^{-6}\rm\ cm^{-3}$, and an intrinsic scatter of $\sigma_p=0.26\pm0.02$ dex.
The derived scale height indicates that the \mgii gas is vertically concentrated, consistent with the dense star-forming disk. The mid-plane density $n_0$ represents the average density within the clouds.

It is important to note that a formal covering fraction ($C_f$) cannot be derived for the stellar sample presented in Table~\ref{tab:mg2stars}. These sightlines are compiled from archival abundance studies (e.g., \cite{jensen2007variation}), which inherently report only positive detections. While this observational selection bias precludes the calculation of a stellar $C_f$, it does not compromise our global density modeling. In our Bayesian MCMC framework, the lack of stellar non-detections is compensated by the 439 rigorous non-detections from the extragalactic sample. Since the total vertical column density ($N_\perp = n_0 h$) is robustly constrained by these extragalactic upper limits, and the scale height ($h$) is anchored by the stellar detections at various $|z|$, this combined approach effectively prevents the mid-plane density ($n_0$) from being artificially inflated by the detection-only stellar sample. This yields a self-consistent description of the Galactic \mgii distribution.

\begin{figure}[H]
\centering
\includegraphics[width=0.9\columnwidth]{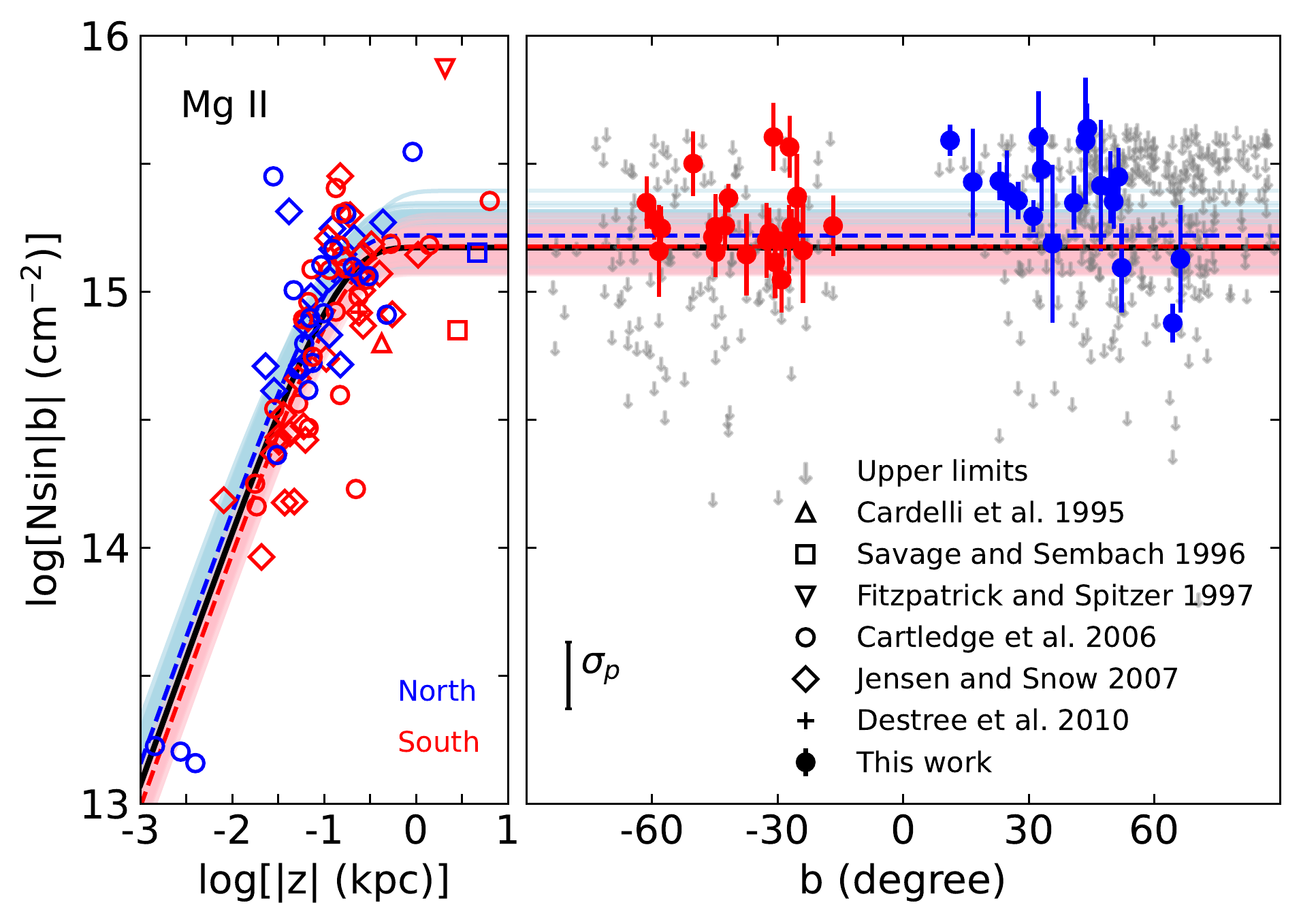}
\caption{Distributions of the projected vertical column density, $\log(N \sin |b|)$.
Left panel: The distribution against vertical height $\log |z|$ for stellar sightlines compiled from literature. Different symbols correspond to references labeled in the lower right.
Right panel: The distribution against Galactic latitude $b$ for extragalactic sightlines from our sample. Solid symbols represent the 43 detections, while gray downward arrows denote the $3\sigma$ upper limits for the 439 non-detections, which are explicitly incorporated into our Bayesian analysis.
The vertical bar in the bottom-left corner illustrates the intrinsic scatter.
In both panels, blue and red colors distinguish \mgii absorbers in the Northern and Southern hemispheres, respectively.
The thick black line represents the best-fit symmetric model ($n_0, h$). The blue and red dashed lines indicate the best-fit models for the Northern ($h_n$) and Southern ($h_s$) hemispheres, respectively. (Note that for extragalactic sources in the right panel, the model predicts a constant $N_\perp = n_0 h$, appearing as horizontal lines).
The thin, semi-transparent blue and red lines display a random subset of models that are drawn from the MCMC posterior distribution, illustrating the uncertainty of the fit.}
\label{fig:stars}
\end{figure}

We display the $\log(N\sin|b|)-\log|z|$ relationships for both stellar and extragalactic sightlines in Figure~\ref{fig:stars}, with the best-fit model indicated by the black line.

The \hi data points generally follow the canonical trend, though with noticeable deviations due to local structures. Unlike the statistically derived parameters for \mgii, the vertical distribution of neutral hydrogen (\hi) has been established through comprehensive absorption analyses of stellar and extragalactic sightlines (e.g., \cite{savage2009extension}). Our \hi sample is inherently biased by \mgii selection and local inhomogeneities. Thus, rather than re-fitting, we adopt the standard Galactic \hi parameters from Savage et al. (2009).

\begin{figure}[H]
\centering
\includegraphics[width=0.9\columnwidth]{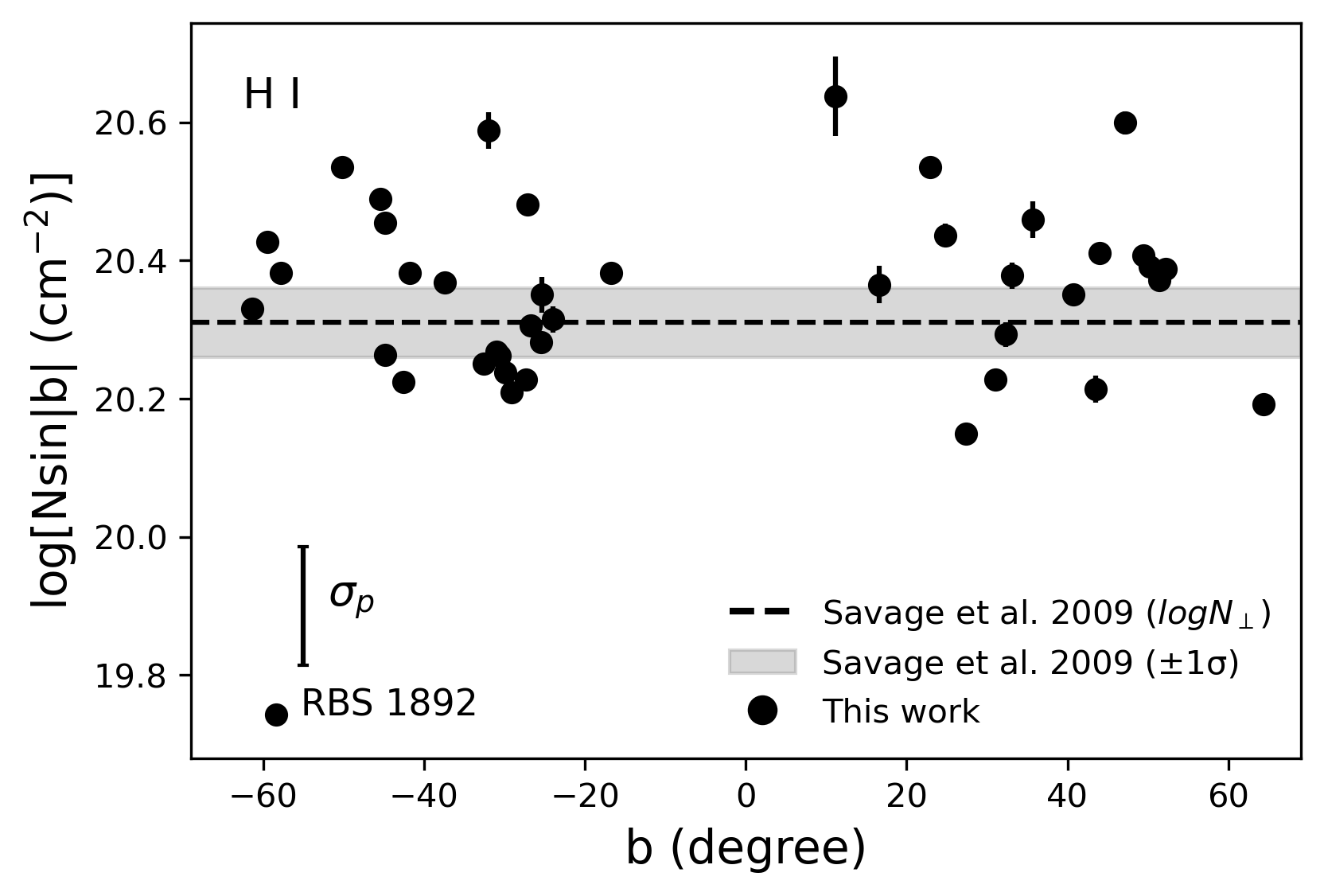}
\caption{Distribution of projected vertical column densities, $\log(N_{\rm HI} \sin |b|)$, for \hi absorbers in our sample plotted against Galactic latitude $b$. The data points represent \hi column densities measured from Ly$\alpha$ absorption. The horizontal dashed line marks the standard Galactic vertical column density of $\log(N_\perp) = 20.31$, with the shaded gray region indicating the $1\sigma$ uncertainty ($\pm 0.05$ dex), adopted from the comprehensive analysis of 139 sightlines by Savage et al. (2009) \cite{savage2009extension}. The vertical bar in the bottom-left corner illustrates the intrinsic scatter ($\sigma_p = 0.172$ dex) of the \hi layer. The sightline toward RBS 1892, which departs significantly from the mean, is labeled.}
\label{fig:h1stars}
\end{figure}

In Figure~\ref{fig:h1stars}, we compare our measurements with the Savage et al. (2009) model ($n_0 \approx 0.276\ \rm cm^{-3}$, $h \approx 0.24\ \rm\ kpc$, $\log N_\perp = 20.31$).
Most data points fall within the expected range. Interestingly, the intrinsic scatter for \mgii derived in this work ($\sigma_p \approx 0.26$ dex) is significantly larger than that reported for \hi ($\sigma_p \approx 0.17$ dex; \cite{savage2009extension}). This indicates that the \mgii gas distribution is spatially more patchy and clumpier than the more ubiquitous neutral hydrogen layer. A notable outlier in the \hi data is the sightline toward RBS 1892 ($\log N_\perp \approx 19.7$), which shows a deficit likely due to the Local Bubble cavity.

\subsection{Asymmetric distribution of cool gas}\label{asym}

Our analysis first examines the spatial correlation between cool gas column density and Galactic latitude. We find Spearman correlation coefficients of $r_s = -0.74 \pm 0.07$ for the northern hemisphere and $r_s = -0.58 \pm 0.10$ for the southern hemisphere. These correlations remain consistent within errors even after excluding the inner Galactic region ($|l| < 45^\circ$), with $r_s = -0.72 \pm 0.10$ (North) and $r_s = -0.58 \pm 0.11$ (South). Regarding Galactic longitude, the correlation appears stronger towards the Galactic center ($r_s = -0.66 \pm 0.07$ for $|l| < 90^\circ$) compared to the anti-center ($r_s = -0.41 \pm 0.19$ for $90^\circ < l < 270^\circ$). However, given the large uncertainty in the anti-center region ($\pm 0.19$), this difference is not statistically significant, suggesting that the vertical slope of the cool gas distribution is relatively consistent across these Galactic longitudes.

While the vertical trends are similar, we observe a significant asymmetry in the total gas abundance. The median logarithmic perpendicular column density in the northern hemisphere is $\log N_\perp=15.38\pm0.04$, which is noticeably higher than that of the southern hemisphere ($\log N_\perp=15.24\pm0.02$). 
These reported uncertainties represent the $1\sigma$ confidence intervals derived from a bootstrap analysis that explicitly incorporates the propagation of individual asymmetric measurement errors.
A two-sample $t$-test on these distributions yields a statistic of $t=3.17$ with a $p$-value of 0.004, indicating a statistically significant difference at the 99\% confidence level. The northern hemisphere hosts a higher concentration of cool gas, with column densities approximately 0.1 dex higher than the south.

To investigate the physical origin of this asymmetry, we first assessed the gas covering fraction ($C_f$) by fitting Equation~\ref{eq:fw} separately for each hemisphere. While the characteristic column density is identical for both ($\log N_c = 15.11\pm0.05$), the normalization constants differ markedly: $C=0.90\pm0.07$ for the south compared to $C=0.42\pm0.04$ for the north. As shown in Figure~\ref{ew_fc_div}(b), the covering fraction is significantly larger in the southern hemisphere. This trend persists even when the region within $45^\circ$ of the Galactic center is excluded. Similarly, Figure~\ref{ew_fc_div}(c) indicates that the covering fraction is relatively higher towards the Galactic center compared to the anti-center, consistent with higher gas densities in the inner Galaxy increasing the detectable incidence rate (e.g., \cite{Cherrey2023MusEGF}).

\begin{figure}[H]
\centering
\includegraphics[width=1.0\columnwidth]{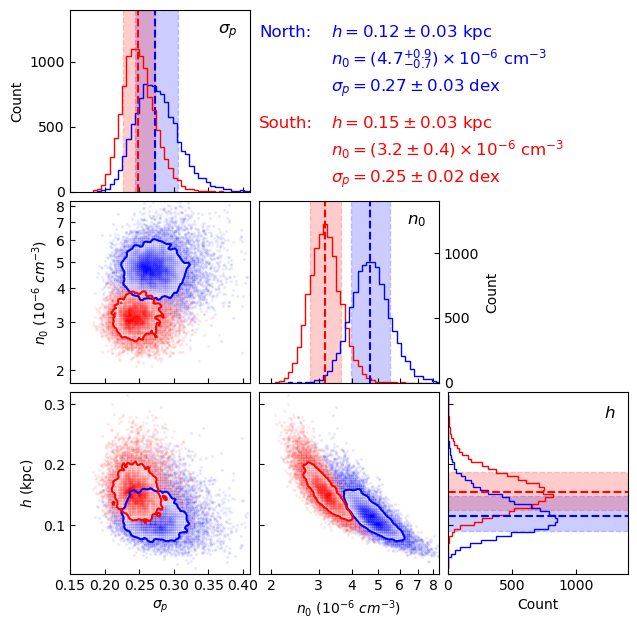}
\caption{Marginalized Bayesian posterior distributions for the model parameters: scale height $h$, mid-plane density $n_0$, and intrinsic scatter $\sigma_p$. The analysis was performed independently for the Galactic northern (blue) and southern (red) hemispheres.
Top and Right panels: The one-dimensional histograms for the intrinsic scatter $\sigma_p$ (top left), mid-plane density $n_0$ (top middle), and scale height $h$ (right). The dashed lines and shaded regions indicate the median values and 68\% confidence intervals, respectively.
Two-dimensional panels: The joint posterior distributions of $n_0$ against $\sigma_p$ (middle left), $h$ against $\sigma_p$ (bottom left), and $h$ against $n_0$ (bottom middle). The contours indicate the 68\% confidence intervals.
The distinct separation in the $n_0$ distributions highlights the systematic difference in mid-plane density between the two hemispheres, while the distributions for $h$ and $\sigma_p$ show significant overlap, indicating consistent vertical extent and scatter.}
\label{fig:level}
\end{figure}

We further performed independent Bayesian fits to model the vertical density profiles, allowing the scale height ($h$) and mid-plane density ($n_0$) to vary for each hemisphere. To decouple intrinsic structure from covering fraction variations, the hemisphere-specific covering fractions ($C_{\rm south}\approx0.90, C_{\rm north}\approx0.42$) were incorporated directly into the likelihood analysis. Figure~\ref{fig:level} presents the marginalized posterior distributions. 

The results reveal that the scale heights are consistent within $1\sigma$ uncertainties ($h_n=0.12\pm0.03\rm\ kpc$ vs. $h_s=0.15\pm0.03\rm\ kpc$). However, the mid-plane volume density in the northern hemisphere ($n_{0,n}=4.7^{+0.9}_{-0.7}\times10^{-6}\rm\ cm^{-3}$) is significantly higher than in the southern hemisphere ($n_{0,s}=3.2\pm0.4\times10^{-6}\rm\ cm^{-3}$), with non-overlapping $1\sigma$ confidence intervals. This implies a physical dichotomy where the northern gas is spatially sparser (lower filling factor) but intrinsically denser compared to the more ubiquitous but diffuse southern gas. 


Notably, adopting a uniform global covering fraction would have biased these estimates, artificially inflating the scale height discrepancy and obscuring this intrinsic density contrast. By explicitly accounting for distinct covering fractions, we effectively removed the bias introduced by the varying cloud incidence rates, revealing that the geometric thickness of the gas disk is likely symmetric ($h_n \approx h_s$) while the asymmetry is primarily driven by the mid-plane density and filling factor.
The intrinsic scatter remains consistent ($\sigma_p\approx0.26$ dex), suggesting similar turbulent properties despite the density contrast.

One potential origin for such a hemispheric discrepancy is the Sun's vertical displacement above the Galactic midplane ($z_\odot$). Various studies using diverse tracers have placed the Sun at $z_\odot \sim 0.015-0.03 \rm\ kpc$ north of the midplane (e.g., \cite{bennett2019vertical}). Given our derived scale height for \mgii\ of $h \approx 0.12-0.15 \rm\ kpc$, the Sun is offset from the center of the gas disk by roughly 15-20\% of the scale height. Geometrically, this northern displacement should result in longer path lengths through the gas layer for sightlines directed toward the south, thereby increasing the observed southern column densities. However, our finding of a higher $N_\perp$ in the North (by $\sim0.14$ dex) is in the opposite direction of this geometric bias. This strongly suggests that the observed northern excess is not a geometric artifact, but an intrinsic structural feature of the Milky Way's cool gas. Interestingly, similar latitude-dependent asymmetries have been reported in high-ionization gas by Zheng et al. (2019) \cite{zheng2019revealing} and Qu et al. (2019) \cite{qu2019warm}. While their physical interpretations vary, the consistent observation of such asymmetries across different ionization states (now including the cool \mgii\ gas) points to a large-scale, intrinsic structural complexity extending from the disk to the halo.

The combination of results (higher mid-plane density $n_0$ yet lower covering fraction $C_f$ in the north) paints a distinct physical picture. As visualized in Figure~\ref{fig:stars}, the cool gas in the northern hemisphere appears to be more clumpy or concentrated in denser structures, whereas the southern gas is more ubiquitously distributed but with a lower average volume density. This structural discrepancy could be related to the Galactic warp, which displaces the effective mid-plane, or to local inhomogeneities in the solar neighborhood.

This north-south asymmetry is corroborated by observations in other wavebands. In the X-ray regime, Snowden et al. (1997) \cite{snowden1997rosat} observed enhanced soft X-ray emission in the northern hemisphere. Similarly, UV observations of warm gas indicate systematically higher column densities for high ions (e.g., \ovi, \civ, and \siiv) along northern sightlines (e.g., \cite{savage2003distribution, wakker2012Characterizing}), likely associated with structures like the North Polar Spur. The coincidence of these asymmetries across the hot, warm, and cool phases suggests a global structural asymmetry in the Milky Way's circumgalactic medium, potentially linked to large-scale feedback processes or the Local Bubble structure.

Furthermore, the stellar distribution shows similar asymmetry. A study by An (2019) \cite{an2019asymmetric} found that the stellar metallicity in the north is significantly higher than in the south within $2 \rm\ kpc$ of the Galactic disk, and recent surveys similarly reveal a vertical asymmetry in [Mg/Fe] at large galactic radii \cite{thomas2024spectroTranslator}.
Since the presence of metals notably enhances radiative cooling efficiency, this metallicity enhancement provides a plausible physical mechanism for the higher cool gas density ($n_0$) and condensation efficiency we observe in the north.

\section{Discussion}
We interpret the \mgii scale height within a simple hydrostatic framework. In this picture, if the vertical distribution were supported solely by thermal pressure, the scale height $h = kT/(\overline{m},g)$ would imply a temperature of $T \sim 10^{4.7}$ K (assuming $\overline{m}=0.73\,m_{\rm H}$ and $g \sim 10^{-8} {\rm cm\ s^{-2}}$). However, this temperature significantly exceeds the typical range where \mgii is abundant in collisional ionization equilibrium ($T_{max} \sim 10^4$ K; e.g., \cite{tumlinson2017circumgalactic}) and would result in the ionization of the vast majority of neutral hydrogen \cite{gnat2007time}. 

Given that \mgii is likely photoionized and exists in cooler gas, this high inferred temperature indicates that thermal pressure alone is insufficient to support the gas. Instead, substantial non-thermal support (e.g., turbulence or magnetic fields) is required to maintain the observed scale height (e.g., \cite{lockman1991vertical}).
This requirement for turbulent support is consistent with the column density distribution analyzed in Section~\ref{sec_comp}, which follows a log-normal-like form typical of a turbulent medium. Furthermore, we must consider that the observed distribution is not solely determined by dynamics; chemical abundance variations also play a role.

\subsection{The effect of Mg depletion on distribution}
Mg depletion regulates how much \mgii remains in the gas phase. Depletion of refractory elements (Mg, Si, Fe) is observed to correlate with hydrogen column density (e.g., \cite{jensen2007variation, Roamn-Duval2021Metal}). We define the depletion of an element X as the difference between its gas-phase and total (dust+gas) abundance, $\log_{10}(X/H)_{\rm gas}-\log_{10}(X/H)_{\rm total}$. Throughout, we adopt a uniform intrinsic (dust+gas) abundance across the \mgii absorbers we analyze. In neutral and mildly photoionized gas, \mgii is the dominant ionization state of Mg; hence, its gas-phase abundance traces Mg depletion.

To compute the \mgii gas abundance, we define the total hydrogen column as $N({\rm H})=N({\rm HI})+2N({\rm H}_2)$, with $N({\rm H_2})$ consistently adopted from \cite{Kalberla2020HI} for all sightlines. For our detection sample, $N({\rm HI})$ is derived from VP fitting. Crucially, the inclusion of 439 non-detections as $3\sigma$ upper limits, with $N({\rm HI})$ sourced from the \hifpi\ survey \cite{bekhti2016hi4pi}, allows for a comprehensive statistical assessment of the depletion trend across the entire observed volume.

\begin{figure}[H]
    \centering
    \includegraphics[width=1.0\columnwidth]{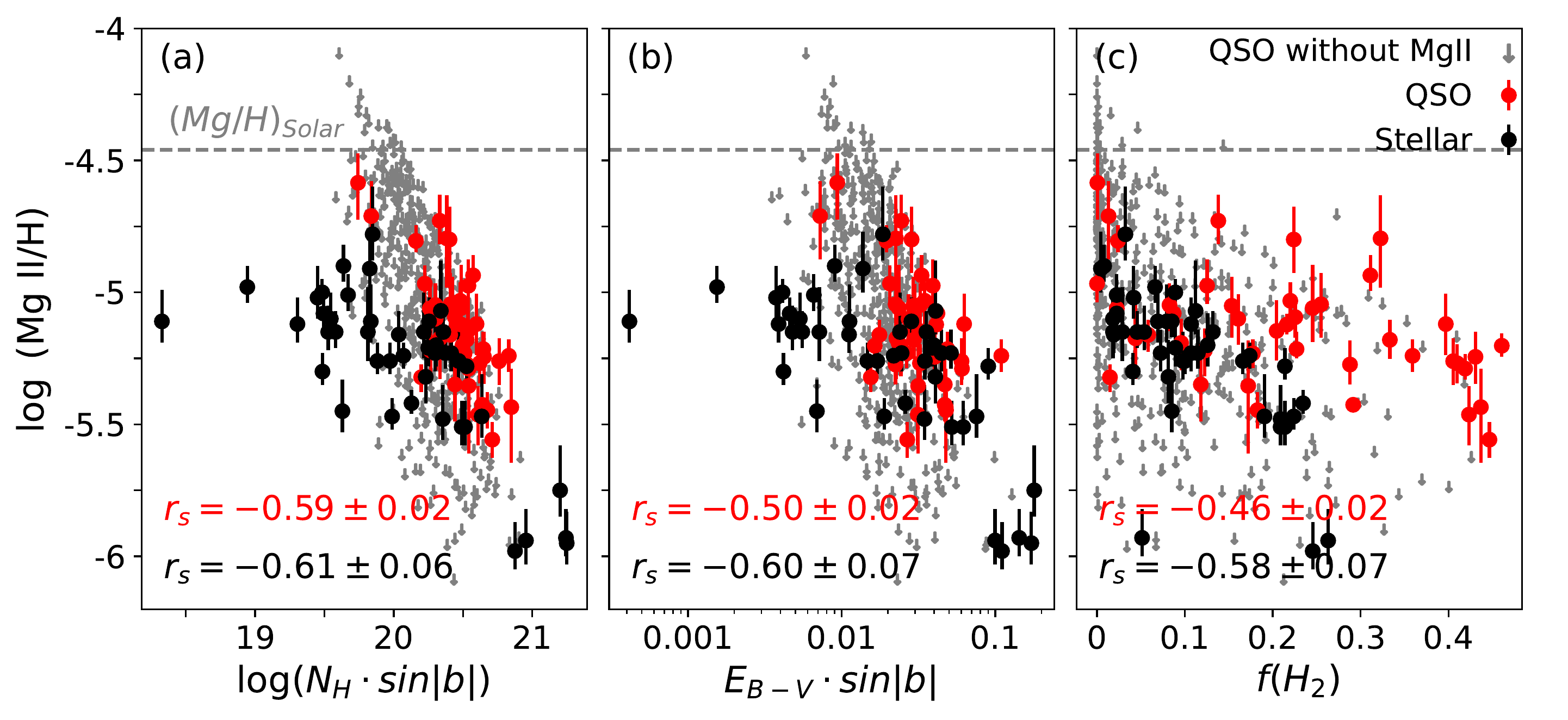}
    \caption{The Mg depletion is plotted against vertical hydrogen density (a), vertical dust extinction (b), and molecular fraction of hydrogen (c). 
    The red points and gray downward arrows represent detections and 3$\sigma$ upper limits for extragalactic sightlines, respectively. The black points denote stellar sightlines from Jensen \& Snow 2007 \cite{jensen2007variation}.
    The horizontal dashed gray line indicates the solar abundance. Note that the x-axes for panel (b) is plotted on a logarithmic scale, representing the geometrically projected linear quantities $E(B-V)\sin|b|$.}
    \label{fig:depletion}
\end{figure}

To account for geometric path-length variations, we adopt the projected vertical column densities. This geometric correction proves particularly essential for sightlines sampling the detected "cloudy" components.
Along stellar sightlines (black points in Figure~\ref{fig:depletion}), using the vertical hydrogen column $N({\rm H})\sin|b|$ significantly removes the bias introduced by slanting sightlines, strengthening the anti-correlation from $r_s=-0.47\pm0.06$ (unprojected, not shown) to $r_s=-0.61\pm0.06$ (Figure~\ref{fig:depletion}a). Similarly, the correlation with reddening in Figure~\ref{fig:depletion}b improves from $r_s=-0.41\pm0.07$ to $r_s=-0.60\pm0.06$ when using $E(B-V)\sin|b|$. The correlation with the molecular fraction (Figure~\ref{fig:depletion}c) remains strong at $r_s=-0.58\pm0.07$.

For the extragalactic sightlines (red points and gray arrows in Figure~\ref{fig:depletion}), the inclusion of upper limits robustly confirms that the negative correlation is a global property of the ISM. When considering the full QSO sample, the anti-correlation with the projected vertical hydrogen density yields $r_s \approx -0.59\pm0.02$ (Figure~\ref{fig:depletion}a). Interestingly, because the inclusion of upper limits drastically expands the dynamic range of the sampled environments, the intrinsic physical depletion trend heavily dominates the overall distribution. As a result, the statistical improvement gained from the $\sin|b|$ projection is less pronounced for the full sample (shifting from an unprojected $r_s \approx -0.63\pm0.02$ to a projected $r_s \approx -0.59$). This indicates that the intrinsic depletion mechanism, rather than geometric projection, is the primary driver of the observed correlation.

Similarly, for dust extinction (Figure~\ref{fig:depletion}b), the correlation with the vertical reddening $E(B-V)\sin|b|$ yields $r_s \approx -0.50\pm0.02$ for the full QSO sample. The correlation with the molecular fraction (Figure~\ref{fig:depletion}c) is $r_s \approx -0.46\pm0.02$. While both the stellar and extragalactic samples exhibit negative correlations, their distributions only converge at the high column density end ($\log N_{\rm H}\sin|b| \gtrsim 20$). At lower column densities, a distinct divergence emerges: for a matched total $N_{\rm H}$ or extinction, the stellar sample exhibits significantly stronger depletion (i.e., lower gas-phase Mg abundance) than the QSO sample.

This divergence perfectly illustrates a unified physical picture of sightline integration, local gas volume density, and the multi-phase ISM. At high $N_{\rm H}$ ($\gtrsim 20$), both stellar and QSO sightlines inevitably intersect massive, dense Cold Neutral Medium (CNM) clouds, where the extreme volume density drives efficient dust condensation, naturally resulting in consistently strong depletion for both samples.

However, below $\log N_{\rm H}\sin|b| \approx 20$, the geometric nature of the sightlines dictates the observed divergence. A critical piece of evidence is the absolute hard cut-off in our extragalactic sample: across all QSO sightlines (including both detections and non-detections), virtually none exhibit a projected column density below $\log N_{\rm H}\sin|b| \approx 19.5$. This establishes a fundamental "baseline" column; a trans-Galactic sightline integrating through the entire halo and disk inevitably accumulates at least $10^{19.5} \rm\ cm^{-2}$ of diffuse, volume-filling gas (predominantly the Warm Neutral Medium, WNM). Consequently, stellar sightlines reporting $N_{\rm H}$ well below 19.5 are not sampling inherently diffuse environments; rather, they are heavily truncated integration paths physically confined within the dense midplane. They accumulate little total $N_{\rm H}$ due to their short lengths, but the local gas they probe is closely associated with dense CNM cores, maintaining the strong depletion observed.

For a full QSO sightline to fall in the intermediate regime ($19.5 \lesssim \log N_{\rm H}\sin|b| \lesssim 20$), it must represent a mixture of phases. The sightlines yielding \mgii non-detections in this regime are those that accumulated the diffuse WNM baseline but geometrically missed the isolated, small-covering-fraction CNM cores entirely. Conversely, the QSO sightlines with \mgii detections in this regime successfully intersected some CNM gas, but their total integrated $N_{\rm H}$ is heavily diluted by the massive WNM baseline. Because the diffuse WNM experiences highly inefficient dust condensation, its dominant presence along the line of sight elevates the overall average gas-phase Mg abundance, naturally causing the QSO sample to exhibit significantly weaker depletion than the purely midplane-confined stellar sightlines at the same total column density.

Finally, we note that while our FUV \mgii detections exclusively trace low-velocity disk gas, the extended extragalactic sightlines inevitably intersect diffuse halo structures such as High-Velocity Clouds (HVCs). Constrained by our sensitivity limits and their extremely low intrinsic volume densities, these halo components naturally fall into our robust sample of non-detections (upper limits). It is crucial to note that HVCs generally possess sub-solar metallicities \cite{wakker2013high}. Because we adopt a uniform solar intrinsic Mg abundance across all sightlines, the absolute metal-poor nature of HVCs artificially lowers their calculated gas-phase abundance, introducing downward scatter among the upper limits at the low-$N_{\rm H}$ end. Physically, however, these dust-poor, diffuse environments lack the necessary conditions for efficient condensation. Their inclusion as non-detections firmly anchors the low-density regime, reinforcing the global trend that the diffuse media dominating these extended sightlines experience fundamentally weaker true depletion.

\subsection{Vertical distribution and Ionization context}

\begin{figure}[H]
\centering
\includegraphics[width=1.0\columnwidth]{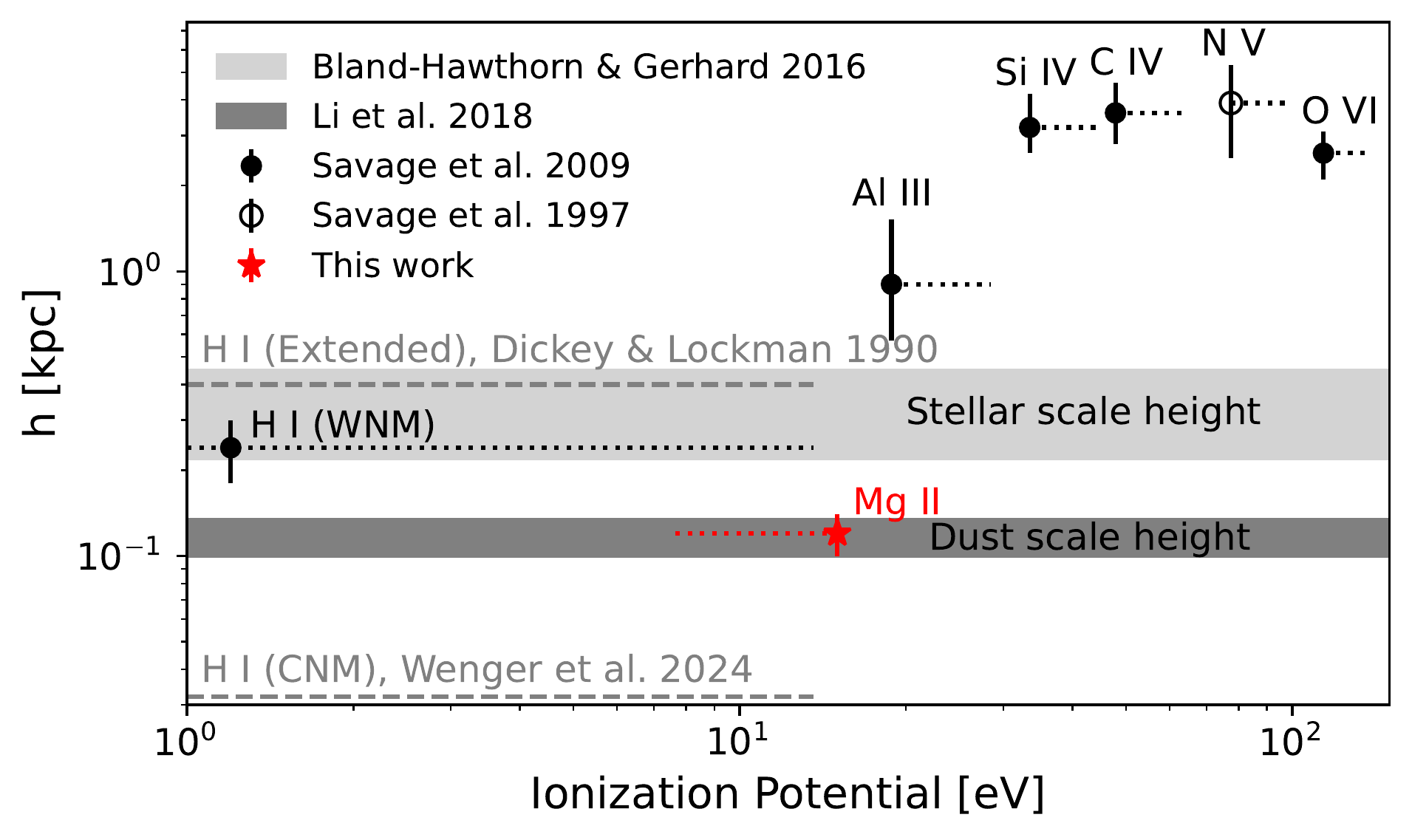}
\caption{The Galactic exponential scale heights of various ions plotted against their ionization potential. The dotted horizontal lines indicate the permissible energy ranges for each ion. Open and filled circles represent results from \cite{savage1997absorption} and \cite{savage2009extension}, respectively. The \mgii scale height determined in this study is marked with a red star ($h \approx 0.12 \rm\ kpc$). For comparison, two gray dashed horizontal lines represent the scale heights of different \hi components: the extended \hi component ($\approx 0.40 \rm\ kpc$, top \cite{dickey1990hi}) and the CNM component ($\approx 0.03 \rm\ kpc$, bottom \cite{wenger2024revisiting}). The light-gray and dark-gray shaded regions denote the scale heights of the stellar old thin disk ($0.22-0.45 \rm\ kpc$; \cite{Bland-Hawthorn2016Galaxy}) and the dust layer of the Milky Way ($0.10-0.13 \rm\ kpc$; \cite{Li2018three}), respectively.}
\label{fig:ionh}
\end{figure}

\mgii primarily traces cool, largely neutral gas within the Galactic disk (where hydrogen is predominantly neutral, though Mg itself is photoionized from \mgi; $n_{\rm H}\!\approx\!10^{-1}\ {\rm cm^{-3}}$, $T\!\lesssim\!10^{4}$ K; e.g., \cite{bergeron1991sample, steidel1994field, churchill2005mgii}).
As illustrated in Figure~\ref{fig:ionh}, the Galactic vertical scale height exhibits a clear correlation with ionization potential: ions with higher ionization energies (e.g., \civ, \ovi) show progressively larger scale heights, extending into the halo. In contrast, low-ionization tracers like \mgii are confined closer to the plane.

Among the ions shown in Figure~\ref{fig:ionh}, the exponential scale height of \mgii is the smallest ($h \approx 0.12 \rm\ kpc$). Quantitatively, this value aligns remarkably well with the scale height of interstellar dust, including the range of $0.10-0.13 \rm\ kpc$ derived by Li et al. (2018) \cite{Li2018three} (dark-gray band) and the scale height of $h \approx 0.152 \rm\ kpc$ reported by Diplas et al. (1994) \cite{diplas1994IUE}.

Comparing our result with the neutral hydrogen (\hi) components reveals the exact multi-phase picture established in Section 4.1. Our \mgii scale height is significantly smaller than the extended \hi component ($h \approx 0.4 \rm\ kpc$; \cite{dickey1990hi}) and the WNM scale height \cite{savage2009extension}. However, its relationship with the CNM depends on the specific component being traced. While historical studies and broad surveys in the solar neighborhood report Gaussian scale heights of $\sigma_z \approx 0.13$-$0.16 \rm\ kpc$ for CNM clouds (e.g., \cite{crovisier1978kinematics, dickey2022gaskap}), Wenger et al. (2024) \cite{wenger2024revisiting} recently corrected for statistical biases and derived a much thinner exponential scale height of $h \approx 0.032 \rm\ kpc$ for the dense CNM cores. Our derived $h \approx 0.12 \rm\ kpc$ is notably larger than this strictly confined core component but falls below the local stellar scale height of the old thin disk ($0.22-0.45 \rm\ kpc$; light-gray band; e.g., \cite{Bland-Hawthorn2016Galaxy}).
We assert that our detected \mgii exclusively traces the cool, neutral gas associated with the CNM phase, and the apparent broadening of its scale height relative to the CNM cores is a direct consequence of dust depletion and geometric selection.

First, as established in Section 4.1 and reflected by the highly depleted stellar sample in Figure~\ref{fig:depletion}, severe condensation of Mg onto dust grains is driven by extreme local volume densities, which peak within the Galactic midplane ($z \approx 0$). This intense midplane depletion effectively suppresses the gas-phase \mgii column density at the very center of the disk, thereby flattening the vertical profile and mathematically broadening the derived exponential scale height of the gas-phase species relative to the underlying total hydrogen mass distribution \cite{savage1996interstellar}.

Second, the vast population of our QSO non-detections at higher latitudes tightly constrains the upper boundary of this vertical distribution. Structurally confined, \mgii-bearing neutral structures, such as CNM cores and their immediate envelopes, possess a limited covering fraction on the sky. Our measured detection rate of $C_f = 0.32$ for $\log N \gtrsim 15$ quantifies this scarcity. Consequently, extragalactic sightlines naturally miss these structures when they pierce regions outside the immediate denser midplane. When a trans-Galactic sightline clears this region ($|z| \gtrsim 0.12 \rm\ kpc$), it predominantly intersects the diffuse WNM or halo gas. The accumulated gas-phase \mgii from this diffuse background naturally falls below our sensitivity threshold, yielding a non-detection. Therefore, rather than being artificially suppressed by our detection limit, the $h \approx 0.12 \rm\ kpc$ scale height robustly characterizes the true vertical extent of the detectable, slightly-less-depleted envelopes of the CNM layer, firmly distinguishing it from the much thicker WNM disk.

Distance-constrained absorption toward halo stars supports this picture, indicating that the bulk of cool material traced by \caii, \nai, and \feii is confined within $\sim 3 \rm\ kpc$ of the disk \cite{bish2019galactic, werk2019nature}, while highly ionized species (\siiv, \civ, \ovi) reveal a significant baryonic mass reservoir extending into the circumgalactic medium \cite{zheng2019revealing, qu2019warm, qu2020circumgalactic}.

\subsection{Milky Way vs. external \mgii}

The Milky Way offers an inside-out view through pencil-beam sightlines that originate in (or cross) the dense mid-plane, whereas extragalactic studies probe external halos in transverse absorption against background QSOs/galaxies. This geometric difference naturally selects two partially distinct \mgii-bearing populations: the Milky Way view is most sensitive to denser, cooler, dust-enriched gas near the disk, whereas the transverse view preferentially intercepts more diffuse, extended halo gas at larger radii and heights. As a consequence, the observed covering fractions and their radial/vertical trends need not match between the two modalities, even for similar galaxy populations.

For clarity, we do not directly compare $C_f$ defined by an NUV equivalent-width threshold in $W_{2796}$ with the FUV-based $C_f$ defined by a \mgii column-density threshold. In our FUV analysis, the measured equivalent width scales approximately linearly with $\log N$, allowing a straightforward mapping from the detection limit in equivalent width to a column-density threshold. In the NUV, however, $W_{2796}$ depends on both $\log N$ and the Doppler parameter $b$, and the literature convention of ``strong'' \mgii refers to $W_{2796}>1\,\text{\AA}$, which is not directly convertible to our FUV thresholds. Therefore, a single function $W(\log N)$ cannot be used to place FUV and NUV \mgii on an identical $C_f$ footing.

Extragalactic measurements further show that the \mgii covering fraction increases toward smaller impact parameters $\rho$ (i.e., closer to galactic centers) and that the central $C_f$ tends to be higher at higher redshift (see their Equation 8; \cite{lan2020coevolution}). Extrapolating their parameterization to $z\!\sim\!0$ and Milky Way-like star-forming hosts yields a transverse $C_f\!\sim\!0.48$ within $\rho\!\lesssim\!10 \rm\ kpc$ for \emph{strong} NUV absorbers ($W_{2796}\!>\!1$\,\AA). Notably, small $\rho$ lines of sight have an increased probability of intersecting the disk or inner, denser layers, making them more akin to Milky Way inside-out geometry. Using the inside-out to transverse mapping of Tumlinson \& Fang 2005 \cite{tumlinson2005hot}, our inside-out measurement $C_f(\log N\!\gtrsim\!15)\!\approx\!0.32$ implies a transverse value $\approx\!\tfrac{4}{3}C_f\!\approx\!0.43$, close to the extrapolated $\sim\!0.48$ at $\rho\!\lesssim\!10 \rm\ kpc$.

While we do not equate the NUV and FUV thresholds directly, this numerical concordance suggests that the strong NUV \mgii population near galaxy centers typically reaches column densities of order $\log N\!\sim\!15$.
This value is consistent with the empirical properties of strong \mgii absorbers ($W_{2796}\!>\!1$ \AA), which are known to trace high column density gas with $\log N_{HI} \sim 19-21$ \cite{rao2006damped,napolitano2025composite}. In such systems, while the \mgii $\lambda2796$ transition is heavily saturated, measurements using weaker lines such as \mgii $\lambda1239$ confirm that the actual column density indeed reaches the $\log N \sim 15$ regime.

The apparent paucity of \mgii detections far from the Milky Way disk does not by itself imply the absence of \mgii. A primary reason is sensitivity: the FUV \mgii doublet has much smaller oscillator strengths than the NUV \mgii doublet (by roughly three orders of magnitude), so at fixed S/N the FUV transition requires substantially higher column densities for detection. This makes our FUV \mgii limits less sensitive than many commonly used ions with stronger $f$-values, and it naturally pushes a large fraction of the Milky Way-halo \mgii reservoir below current FUV detection thresholds. Sparse angular sampling at high latitudes further compounds this effect by increasing the chance of missing patchy structures.

A unified picture thus emerges. Both the Milky Way and external galaxies likely host extended \mgii-bearing halos. However, the observed distributions are significantly shaped by the interplay between physical conditions and detectability thresholds. In the Milky Way, the limited sensitivity of the weak FUV transitions biases our detection toward the dense, neutral gas (CNM) concentrated near the plane, resulting in one of the smallest scale heights among low ions. In contrast, transverse sightlines through external halos are often capable of detecting lower-density, more diffuse \mgii\ gas extending to larger radii, highlighting a component that likely exists in the Milky Way but remains below the detection limit of our current sample.

\section{Summary and Conclusions}
In this study, we characterized the \mgii column density distribution around the Galactic disk by analyzing 43 low-velocity ($|v|<40\ {\rm km\ s^{-1}}$) absorbers across 482 extragalactic sightlines. By integrating stellar sightline data and employing Bayesian inference, we provide a comprehensive map of the \mgii gas phase. Our key findings are as follows:

1. Distribution and Phase Transition: The \mgii covering fraction ($C_f$) for $\log N_{\rm MgII} > 15$ is $32\pm5\%$. Notably, the detection rate as a function of the equivalent width threshold ($W_{th}$) is best characterized by an exponential decay model. This distribution reveals a transition from a pervasive, diffuse medium (following a cosmic power-law) to localized, dense structures. The observed steep decline at high column densities (deviating from the power-law) likely reflects the saturation of the turbulent log-normal spectrum combined with gas-phase magnesium depletion within the dense ISM.

2. Disk Morphology and Vertical Structure: We found a strong anti-correlation ($r_s = -0.62 \pm 0.06$) between column density and Galactic latitude. Bayesian modeling, accounting for censored data and intrinsic scatter, constrained the disk parameters to a projected perpendicular column density $\log N_\perp=15.29^{+0.04}_{-0.03}$, a scale height $h = 0.12\pm0.02\ \rm\ kpc$, and a mid-plane density $n_0 = (3.9\pm0.4)\times 10^{-6}\ \rm cm^{-3}$.  These results indicate that \mgii gas is tightly confined to the star-forming disk.

3. North-South Asymmetry: A pronounced north-south asymmetry exists in the \mgii distribution. While the southern hemisphere has a higher covering fraction ($C_f \approx 0.90$ vs. $0.41$), the northern hemisphere exhibits a significantly higher mean abundance ($\log N_\perp=15.38$ vs. $15.24$; $p=0.004$). Independent Bayesian fits reveal that while the vertical scale heights are consistent, the north possesses a significantly higher mid-plane volume density ($n_{0,n} \approx 4.7 \times 10^{-6}\ \rm cm^{-3}$ vs. $n_{0,s} \approx 3.2 \times 10^{-6}\ \rm cm^{-3}$). This suggests that the northern cool gas is more spatially concentrated and clumpy, whereas the southern gas is more diffuse and ubiquitously distributed.

4. Volume Density Dependence and Line-of-Sight Integration in Mg Depletion: Magnesium depletion displays a robust anti-correlation with the projected vertical hydrogen column density, interstellar extinction, and molecular hydrogen fraction. Rather than a simple uniform depletion law, our results reveal a multi-phase physical picture driven by local volume density and geometric integration. At high column densities ($\log N_{\rm H}\sin|b| \gtrsim 20$), extragalactic and stellar sightlines seamlessly converge, as both inherently intersect massive, dense CNM clouds where extreme volume density drives efficient dust condensation. However, a stark divergence emerges at lower column densities due to distinct line-of-sight geometries. We identify an absolute lower bound of $\log N_{\rm H}\sin|b| \approx 19.5$ across our entire extragalactic sample, establishing a minimum baseline column composed of diffuse, volume-filling gas (predominantly WNM). Consequently, stellar sightlines yielding low $N_{\rm H}$ are strictly truncated paths confined to the dense midplane, maintaining strong depletion. In contrast, QSO sightlines in the intermediate regime ($19.5 \lesssim \log N_{\rm H}\sin|b| \lesssim 20.0$) represent full trans-Galactic integrations dominated by this diffuse WNM baseline. Within this regime, non-detections geometrically miss the structurally confined CNM cores and their immediate envelopes entirely. Conversely, \mgii detections do intersect some denser gas, but their integrated depletion signature is heavily diluted by the massive, weakly depleted WNM background. Both scenarios naturally result in the systematically weaker average depletion observed in the extragalactic sample. Furthermore, this density-driven picture naturally accommodates diffuse halo structures such as High-Velocity Clouds (HVCs). While our sensitivity threshold generally precludes individual detections of such low-density, sub-solar metallicity gas, their physical properties align with the extreme low-depletion end of this continuum. Sightlines intersecting these diffuse halo structures naturally populate our robust pool of non-detections at lower projected column densities, anchoring the global depletion trend without altering the bulk mass budget of the disk.

5. Multi-Phase Vertical Structure and Geometric Selection: We derive an exponential scale height of $h \approx 0.12 \rm\ kpc$ for the \mgii-bearing gas, which physically aligns with the vertical extent of the Galactic dust layer. Crucially, this vertical distribution sits perfectly between the strictly confined CNM cores ($h \approx 0.032 \rm\ kpc$) and the extended WNM ($h \approx 0.4 \rm\ kpc$), reflecting the exact multi-phase depletion physics established above. We attribute this intermediate scale height to extreme dust depletion at the very center of the midplane ($z \approx 0$), which severely suppresses gas-phase \mgii and effectively broadens the vertical profile relative to the densest intrinsic CNM mass distribution. Furthermore, the robust pool of QSO non-detections tightly locks down the upper boundary of this distribution. It proves that once trans-Galactic sightlines clear the denser midplane, they merely accumulate diffuse WNM and halo gas, with the resulting gas-phase \mgii naturally falling below our detection sensitivity. This geometric selection is mathematically quantified by our measured covering fraction of $C_f = 0.32$ for $\log N \gtrsim 15$, confirming that extended extragalactic surveys inherently miss these structurally confined, \mgii-bearing neutral envelopes unless piercing close to the midplane, thereby robustly defining the true vertical boundary of the detectable cool gas.

6. Connection to Transverse Observations: Our inside-out $C_f$ measurements, when mapped to a transverse geometry, are broadly consistent with the extrapolated results for strong \mgii absorbers ($W_{2796} > 1$ \AA) observed in extragalactic surveys (e.g., Lan et al. 2020). This numerical comparability suggests that the strong \mgii populations in the inner regions of star-forming galaxies typically reach column densities of $\log N \sim 15$. This inference is physically supported by our detection of the weak \mgii $\lambda1239$ transition and aligns with the neutral hydrogen column densities ($\log N_{HI} \sim 19-21$) characteristic of Damped Lyman-$\alpha$ systems. Such a link reinforces the consistency between the local ISM and the absorbers detected in the spectra of distant galaxies.

\Acknowledgements{
We are deeply indebted to Zheng Yong for her meticulous contributions to grammatical refinement and data processing, which significantly enhanced the clarity and accuracy of this work.
This work is supported by the National Natural Science Foundation of China (NSFC) under Nos. 11890692, 12133008, 12221003, 12373007, 12422302. We acknowledge the science research grants from the China Manned Space Project with No. CMS-CSST-2025-A04 and No. CMS-CSST-2025-A10. X.J. acknowledges the support from grant No. JAT241087, provided by the Fujian Provincial Department of Education.
This research has made use of the \hsla database, developed and maintained at STScI, Baltimore, USA.
}

\begin{table*}
\centering
\footnotesize
\begin{threeparttable}
\caption{The parameters of the \mgii absorbers}
\begin{tabular}{lrrcccccrr}
\toprule
Target Name &       $\log\ N$\tnote{1)} &                         $\log N$\tnote{2)} &                $b_D$\tnote{2)} &                       $v$\tnote{2)} & $\mathcal{W}_{tot}$\tnote{3)} &    SNR \\
& ($\rm cm^{-2}$) & ($\rm cm^{-2}$) & ($\rm km\ s^{-1}$) & ($\rm km\ s^{-1}$) & (m\AA) & \\ 
\midrule
NGC-3783 & $15.84_{- 0.04}^{+ 0.07}$ & $15.82 \pm { 0.03}$ & $ 13.3 \pm {  2.2}$ & $ -8.3 \pm {  0.6}$ & $ 74.5 \pm {  2.3}$ & 93.5 \\
3C273 & $14.92_{- 0.06}^{+ 0.05}$ & $15.01 \pm { 0.32}$ & $  8.8 \pm { 21.8}$ & $-11.7 \pm {  5.3}$ & $ 10.8 \pm {  1.2}$ & 113.2 \\
3C066A & $15.80_{- 0.08}^{+ 0.09}$ & $15.78 \pm { 0.05}$ & $ 12.7 \pm {  3.1}$ & $ -9.1 \pm {  0.9}$ & $ 68.0 \pm {  8.6}$ & 25.9 \\
RBS1666 & $15.89_{- 0.09}^{+ 0.10}$ & $15.83 \pm { 0.08}$ & $ 24.3 \pm {  7.7}$ & $-11.0 \pm {  2.5}$ & $ 80.4 \pm { 11.2}$ & 20.0 \\
PKS0637-752 & $15.91_{- 0.07}^{+ 0.10}$ & $15.87 \pm { 0.06}$ & $ 15.2 \pm {  4.5}$ & $ -8.9 \pm {  1.3}$ & $ 82.2 \pm {  8.8}$ & 24.3 \\
FAIRALL9 & $15.32_{- 0.07}^{+ 0.06}$ & $15.36 \pm { 0.08}$ & $  9.4 \pm {  5.1}$ & $ -9.0 \pm {  1.3}$ & $ 25.8 \pm {  3.4}$ & 49.1 \\
RXSJ00437+3725 & $15.73_{- 0.13}^{+ 0.12}$ & $15.68 \pm { 0.09}$ & $ 27.8 \pm {  9.5}$ & $-27.5 \pm {  3.2}$ & $ 59.4 \pm { 12.7}$ & 18.2 \\
ESO462-G09 & $15.51_{- 0.07}^{+ 0.07}$ & $15.52 \pm { 2.07}$ & $ 14.8 \pm {152.1}$ & $  6.6 \pm { 43.7}$ & $ 38.6 \pm {  5.1}$ & 23.3 \\
SDSSJ080908.13+461925.6 & $15.88_{- 0.13}^{+ 0.13}$ & $15.82 \pm { 0.42}$ & $ 21.8 \pm { 37.0}$ & $ -6.0 \pm { 11.8}$ & $ 77.7 \pm { 14.5}$ & 14.9 \\
H1821+643 & $15.69_{- 0.04}^{+ 0.06}$ & $15.66 \pm { 0.19}$ & $ 22.0 \pm { 17.0}$ & $-15.3 \pm {  5.4}$ & $ 56.0 \pm {  3.7}$ & 61.1 \\
Q1545+210 & $15.53_{- 0.10}^{+ 0.10}$ & $15.52 \pm { 0.12}$ & $ 19.2 \pm { 10.0}$ & $-13.4 \pm {  3.1}$ & $ 39.7 \pm {  7.4}$ & 27.7 \\
SDSSJ094733.21+100508.7 & $15.75_{- 0.19}^{+ 0.16}$ & $15.65 \pm { 0.13}$ & $ 11.2 \pm {  8.5}$ & $  2.5 \pm {  2.3}$ & $ 62.3 \pm { 19.5}$ & 11.3 \\
1H-2129-624 & $15.43_{- 0.07}^{+ 0.07}$ & $15.45 \pm { 0.08}$ & $ 14.5 \pm {  6.1}$ & $ -2.8 \pm {  1.8}$ & $ 32.8 \pm {  4.6}$ & 40.9 \\
NGC-985 & $15.34_{- 0.06}^{+ 0.05}$ & $15.39 \pm { 0.05}$ & $  8.7 \pm {  3.6}$ & $-10.8 \pm {  0.9}$ & $ 27.4 \pm {  2.8}$ & 61.2 \\
PMNJ1103-2329 & $15.74_{- 0.12}^{+ 0.12}$ & $15.71 \pm { 0.09}$ & $ 16.8 \pm {  6.7}$ & $-12.1 \pm {  2.0}$ & $ 61.1 \pm { 12.4}$ & 17.3 \\
PKS2005-489 & $15.47_{- 0.11}^{+ 0.09}$ & $15.53 \pm { 0.53}$ & $  6.4 \pm { 30.7}$ & $  1.4 \pm {  6.8}$ & $ 35.5 \pm {  7.2}$ & 24.2 \\
PDS456 & $16.55_{- 0.21}^{+ 0.79}$ & $16.30 \pm { 0.06}$ & $ 21.0 \pm {  4.7}$ & $ -1.9 \pm {  1.6}$ & $198.2 \pm { 28.0}$ & 8.2 \\
LBQS-1435-0134 & $15.56_{- 0.09}^{+ 0.07}$ & $15.53 \pm { 0.11}$ & $ 25.5 \pm { 10.9}$ & $  0.8 \pm {  3.6}$ & $ 42.0 \pm {  6.5}$ & 33.8 \\
PKS1101-325 & $15.77_{- 0.11}^{+ 0.12}$ & $15.83 \pm { 0.10}$ & $  6.9 \pm {  4.2}$ & $-10.3 \pm {  1.0}$ & $ 64.4 \pm { 12.2}$ & 17.1 \\
SBS1415+437-OBJECT-1 & $15.17_{- 0.16}^{+ 0.13}$ & $15.24 \pm { 0.14}$ & $  8.3 \pm {  9.9}$ & $ 13.9 \pm {  2.3}$ & $ 18.8 \pm {  5.8}$ & 27.0 \\
MR2251-178 & $15.40_{- 0.08}^{+ 0.07}$ & $15.30 \pm { 0.09}$ & $ 13.6 \pm {  6.7}$ & $ -8.9 \pm {  1.9}$ & $ 31.1 \pm {  4.5}$ & 42.4 \\
MRK1513 & $15.36_{- 0.10}^{+ 0.08}$ & $15.48 \pm { 0.12}$ & $  3.6 \pm {  2.0}$ & $  0.8 \pm {  1.4}$ & $ 28.2 \pm {  5.1}$ & 32.2 \\
1ES1553+113 & $15.80_{- 0.06}^{+ 0.08}$ & $15.75 \pm { 0.09}$ & $ 22.8 \pm {  7.9}$ & $  1.3 \pm {  2.6}$ & $ 68.0 \pm {  5.8}$ & 37.7 \\
MARK509 & $15.50_{- 0.02}^{+ 0.03}$ & $15.51 \pm { 0.08}$ & $ 15.9 \pm {  6.3}$ & $ -1.8 \pm {  1.9}$ & $ 38.2 \pm {  1.2}$ & 163.0 \\
PHL1811 & $15.31_{- 0.07}^{+ 0.07}$ & $15.40 \pm { 0.33}$ & $  4.0 \pm {  9.5}$ & $ -8.6 \pm {  4.5}$ & $ 25.0 \pm {  3.5}$ & 46.2 \\
PG0804+761 & $15.58_{- 0.04}^{+ 0.05}$ & $15.56 \pm { 0.44}$ & $ 23.0 \pm { 39.8}$ & $-10.0 \pm { 12.9}$ & $ 44.7 \pm {  3.1}$ & 68.4 \\
ESO265-G23 & $15.97_{- 0.13}^{+ 0.16}$ & $15.94 \pm { 0.08}$ & $ 13.6 \pm {  5.2}$ & $ -8.9 \pm {  1.6}$ & $ 92.9 \pm { 18.1}$ & 12.4 \\
RXSJ01188+3836 & $15.55_{- 0.16}^{+ 0.14}$ & $15.60 \pm { 0.13}$ & $  8.1 \pm {  8.4}$ & $ -0.9 \pm {  2.0}$ & $ 42.2 \pm { 11.8}$ & 15.7 \\
PG1011-040 & $15.53_{- 0.07}^{+ 0.08}$ & $15.55 \pm { 0.13}$ & $ 13.7 \pm {  9.1}$ & $ -5.7 \pm {  2.6}$ & $ 41.0 \pm {  5.8}$ & 33.0 \\
PKS0405-123 & $15.54_{- 0.04}^{+ 0.04}$ & $15.53 \pm { 0.05}$ & $ 19.3 \pm {  4.0}$ & $ -5.4 \pm {  1.3}$ & $ 41.1 \pm {  2.3}$ & 88.6 \\
NGC-7469 & $15.36_{- 0.02}^{+ 0.02}$ & $15.38 \pm { 0.04}$ & $ 15.1 \pm {  2.7}$ & $ -3.3 \pm {  0.8}$ & $ 28.3 \pm {  1.1}$ & 166.6 \\
RXS-J23218-7026 & $15.40_{- 0.10}^{+ 0.08}$ & $15.43 \pm { 0.14}$ & $ 12.9 \pm {  9.9}$ & $ -7.9 \pm {  2.8}$ & $ 30.9 \pm {  5.5}$ & 33.5 \\
IZW1 & $15.61_{- 0.09}^{+ 0.09}$ & $15.76 \pm { 0.22}$ & $  4.0 \pm {  2.9}$ & $-12.5 \pm {  0.8}$ & $ 47.3 \pm {  7.3}$ & 26.2 \\
PKS1302-102 & $15.20_{- 0.14}^{+ 0.10}$ & $15.37 \pm { 0.14}$ & $ 11.3 \pm {  9.6}$ & $-14.8 \pm {  2.6}$ & $ 19.6 \pm {  5.3}$ & 29.7 \\
PG0052+251 & $15.36_{- 0.12}^{+ 0.11}$ & $15.43 \pm { 0.09}$ & $  5.9 \pm {  6.8}$ & $-11.1 \pm {  1.2}$ & $ 28.3 \pm {  6.4}$ & 26.6 \\
RBS1892 & $15.23_{- 0.14}^{+ 0.11}$ & $15.25 \pm { 0.20}$ & $ 21.0 \pm { 17.6}$ & $ -4.4 \pm {  5.5}$ & $ 21.0 \pm {  5.5}$ & 33.2 \\
PKS0552-640 & $15.41_{- 0.10}^{+ 0.09}$ & $15.44 \pm { 0.72}$ & $ 12.5 \pm { 50.9}$ & $  2.4 \pm { 14.0}$ & $ 31.6 \pm {  6.2}$ & 29.7 \\
IRAS-L06229-6434 & $15.54_{- 0.10}^{+ 0.09}$ & $15.54 \pm { 0.08}$ & $ 17.9 \pm {  6.4}$ & $ -2.3 \pm {  2.0}$ & $ 42.0 \pm {  7.4}$ & 28.6 \\
MRK1392 & $15.47_{- 0.08}^{+ 0.07}$ & $15.48 \pm { 0.08}$ & $ 17.0 \pm {  6.5}$ & $-10.6 \pm {  2.0}$ & $ 35.4 \pm {  5.0}$ & 38.3 \\
AKN-564 & $15.74_{- 0.12}^{+ 0.11}$ & $15.70 \pm { 0.12}$ & $ 19.2 \pm {  9.7}$ & $ -8.9 \pm {  3.0}$ & $ 60.6 \pm { 12.1}$ & 17.1 \\
ESO-141-55 & $15.60_{- 0.04}^{+ 0.06}$ & $15.62 \pm { 0.06}$ & $  9.6 \pm {  4.0}$ & $ -0.4 \pm {  1.0}$ & $ 46.8 \pm {  3.7}$ & 53.5 \\

\bottomrule
\end{tabular}
\label{lqso}
\begin{tablenotes}
\item[1)] The parameters obtained by the COG.
\item[2)] The parameters obtained by the VP.
\item[3)] The combined equivalent width of the doublet ($W_{1239} + W_{1240}$)
\end{tablenotes}
\end{threeparttable}
\end{table*}

\begin{table*}
\centering
\footnotesize
\caption{The parameters of the \hi\ absorbers}
\begin{tabular}{lcccccc}
\toprule
Target Name & b & l & E(B-V) & $\log\ N(\rm HI)$ & $\log\ N(\rm HI)_{\rm HI4PI}$ & $\log\ N(\rm H_2)$ \\
 & (degree) & (degree) & (mag) & ($\rm cm^{-2}$) & ($\rm cm^{-2}$) & ($\rm cm^{-2}$) \\
\midrule
               NGC-3783 &  22.948 & 287.456 &   0.13 & $20.94 \pm { 0.01}$ & 20.95 &    20.11 \\
                  3C273 &  64.360 & 289.951 &   0.02 & $20.24 \pm { 0.01}$ & 20.16 &    18.11 \\
                 3C066A & -16.767 & 140.143 &   0.08 & $20.92 \pm { 0.01}$ & 20.90 &    20.23 \\
                RBS1666 & -31.003 & 358.733 &   0.05 & $20.56 \pm { 0.01}$ & 20.67 &    19.46 \\
            PKS0637-752 & -27.158 & 286.368 &   0.09 & $20.82 \pm { 0.01}$ & 20.84 &    19.68 \\
               FAIRALL9 & -57.827 & 295.073 &   0.02 & $20.45 \pm { 0.01}$ & 20.27 &    18.94 \\
         RXSJ00437+3725 & -25.424 & 121.233 &   0.05 & $20.65 \pm { 0.01}$ & 20.70 &    19.88 \\
             ESO462-G09 & -31.949 &  11.326 &   0.09 & $20.86 \pm { 0.03}$ & 20.70 &    19.91 \\
SDSSJ080908.13+461925.6 &  32.289 & 173.322 &   0.05 & $20.57 \pm { 0.02}$ & 20.54 &    19.72 \\
              H1821+643 &  27.417 &  94.002 &   0.04 & $20.49 \pm { 0.01}$ & 20.53 &    18.57 \\
              Q1545+210 &  49.458 &  33.898 &   0.05 & $20.53 \pm { 0.01}$ & 20.52 &    20.10 \\
SDSSJ094733.21+100508.7 &  43.539 & 225.372 &   0.03 & $20.38 \pm { 0.02}$ & 20.36 &    19.75 \\
            1H-2129-624 & -42.523 & 331.143 &   0.03 & $20.39 \pm { 0.01}$ & 20.44 &    15.52 \\
                NGC-985 & -59.490 & 180.837 &   0.03 & $20.49 \pm { 0.01}$ & 20.55 &    19.52 \\
          PMNJ1103-2329 &  33.080 & 273.190 &   0.11 & $20.64 \pm { 0.02}$ & 20.71 &    20.16 \\
            PKS2005-489 & -32.601 & 350.373 &   0.06 & $20.52 \pm { 0.01}$ & 20.56 &    18.58 \\
                 PDS456 &  11.164 &  10.392 &   0.56 & $21.35 \pm { 0.06}$ & 21.29 &    20.80 \\
         LBQS-1435-0134 &  51.375 & 348.718 &   0.04 & $20.48 \pm { 0.01}$ & 20.48 &    19.63 \\
            PKS1101-325 &  24.765 & 278.117 &   0.08 & $20.81 \pm { 0.02}$ & 20.83 &    19.92 \\
   SBS1415+437-OBJECT-1 &  66.199 &  81.957 &   0.01 & $21.08 \pm { 0.01}$ & 19.87 &    17.70 \\
             MR2251-178 & -61.326 &  46.198 &   0.03 & $20.39 \pm { 0.01}$ & 20.42 &    19.55 \\
                MRK1513 & -29.070 &  63.670 &   0.05 & $20.52 \pm { 0.01}$ & 20.56 &    19.37 \\
            1ES1553+113 &  43.964 &  21.909 &   0.05 & $20.57 \pm { 0.01}$ & 20.55 &    19.92 \\
                MARK509 & -29.856 &  35.971 &   0.08 & $20.54 \pm { 0.01}$ & 20.60 &    19.22 \\
                PHL1811 & -44.815 &  47.473 &   0.04 & $20.61 \pm { 0.01}$ & 20.62 &    20.21 \\
             PG0804+761 &  31.033 & 138.279 &   0.03 & $20.52 \pm { 0.01}$ & 20.53 &    20.15 \\
             ESO265-G23 &  16.592 & 285.910 &   0.08 & $20.91 \pm { 0.03}$ & 20.84 &    20.12 \\
         RXSJ01188+3836 & -23.955 & 128.783 &   0.06 & $20.71 \pm { 0.02}$ & 20.67 &    19.06 \\
            HE0226-4110 & -65.775 & 253.941 &   0.02 & $20.10 \pm { 0.01}$ & 20.15 &    17.32 \\
             PG1011-040 &  40.749 & 246.501 &   0.05 & $20.54 \pm { 0.01}$ & 20.55 &    19.93 \\
            PKS0405-123 & -41.756 & 204.927 &   0.06 & $20.56 \pm { 0.01}$ & 20.55 &    19.70 \\
               NGC-7469 & -45.467 &  83.098 &   0.07 & $20.64 \pm { 0.01}$ & 20.65 &    19.95 \\
                PHL1226 & -54.621 & 150.796 &   0.05 & $20.66 \pm { 0.01}$ & 20.59 &    19.80 \\
        RXS-J23218-7026 & -44.837 & 313.292 &   0.05 & $20.42 \pm { 0.01}$ & 20.47 &    19.07 \\
                   IZW1 & -50.175 & 123.748 &   0.08 & $20.65 \pm { 0.01}$ & 20.66 &    20.18 \\
            PKS1302-102 &  52.161 & 308.591 &   0.06 & $20.49 \pm { 0.01}$ & 20.49 &    19.32 \\
             PG0052+251 & -37.438 & 123.908 &   0.05 & $20.58 \pm { 0.01}$ & 20.59 &    20.15 \\
                RBS1892 & -58.367 & 345.898 &   0.01 & $19.81 \pm { 0.01}$ & 19.91 &    16.13 \\
            PKS0552-640 & -30.611 & 273.466 &   0.05 & $20.55 \pm { 0.01}$ & 20.55 &    19.28 \\
       IRAS-L06229-6434 & -27.319 & 274.312 &   0.06 & $20.57 \pm { 0.01}$ & 20.60 &    19.55 \\
                MRK1392 &  50.264 &   2.754 &   0.04 & $20.50 \pm { 0.01}$ & 20.54 &    20.05 \\
                AKN-564 & -25.337 &  92.139 &   0.07 & $20.72 \pm { 0.03}$ & 20.69 &    19.68 \\
             ESO-141-55 & -26.711 & 338.183 &   0.13 & $20.65 \pm { 0.01}$ & 20.69 &    20.21 \\
\bottomrule
\end{tabular}
\label{hi}
\end{table*}

\begin{table*}
\centering
\footnotesize
\begin{threeparttable}
\caption{The \mgii absorbers along the sightlines of stars in the Milky Way.}
\begin{tabular}{lcccccc|lcccccc}
\toprule
       Name &      l &      b &     D\tnote{1)} &    Z\tnote{2)} &              log(N) & Ref &         Name &      l &      b &     D\tnote{1)} &    Z\tnote{2)} &              log(N) & Ref \\
 & degree & degree & pc & pc & & & & degree & degree & pc & pc & & \\
\midrule
BD +35 4258 &  77.19 &  -4.74 &  3100 &  256 & $16.15 \pm { 0.10}$ &   \cite{jensen2007variation} & CPD -59 2603 & 287.59 &  -0.69 &  2630 &   31 & $16.35 \pm { 0.04}$ &   \cite{jensen2007variation} \\
   HD 12323 & 132.91 &  -5.87 &  3900 &  398 & $16.06 \pm { 0.06}$ &   \cite{jensen2007variation} &     HD 13745 & 134.58 &  -4.96 &  1900 &  164 & $16.18 \pm { 0.09}$ &   \cite{jensen2007variation} \\
   HD 15137 & 137.46 &   7.58 &  3300 &  435 & $16.15 \pm { 0.03}$ &   \cite{jensen2007variation} &     HD 27778 & 172.76 & -17.39 &   223 &   66 & $15.42 \pm { 0.08}$ &   \cite{jensen2007variation} \\
   HD 37021 & 209.01 & -19.38 &   450 &  149 & $15.93 \pm { 0.05}$ &   \cite{jensen2007variation} &     HD 37061 & 208.92 & -19.27 &   580 &  191 & $15.78 \pm { 0.04}$ &   \cite{jensen2007variation} \\
   HD 37903 & 206.85 & -16.54 &   910 &  259 & $15.55 \pm { 0.11}$ &   \cite{jensen2007variation} &     HD 40893 & 180.09 &   4.34 &  2800 &  211 & $16.33 \pm { 0.04}$ &   \cite{jensen2007variation} \\
   HD 69106 & 254.52 &  -1.33 &  1600 &   37 & $15.81 \pm { 0.04}$ &   \cite{jensen2007variation} &     HD 91597 & 286.86 &  -2.37 &  6400 &  264 & $16.25 \pm { 0.07}$ &   \cite{jensen2007variation} \\
   HD 91651 & 286.55 &  -1.72 &  3500 &  105 & $16.26 \pm { 0.05}$ &   \cite{jensen2007variation} &     HD 92554 & 287.60 &  -2.02 &  6795 &  239 & $16.37 \pm { 0.06}$ &   \cite{jensen2007variation} \\
   HD 93205 & 287.57 &  -0.71 &  2600 &   32 & $16.32 \pm { 0.04}$ &   \cite{jensen2007variation} &     HD 93222 & 287.74 &  -1.02 &  2900 &   51 & $16.41 \pm { 0.02}$ &   \cite{jensen2007variation} \\
   HD 93843 & 228.24 &  -0.90 &  2700 &   42 & $16.25 \pm { 0.03}$ &   \cite{jensen2007variation} &     HD 94493 & 289.01 &  -1.18 &  2900 &   59 & $16.16 \pm { 0.03}$ &   \cite{jensen2007variation} \\
   HD 99857 & 294.78 &  -4.94 &  3058 &  263 & $16.21 \pm { 0.04}$ &   \cite{jensen2007variation} &     HD 99890 & 291.75 &   4.43 &  3070 &  237 & $16.18 \pm { 0.04}$ &   \cite{jensen2007variation} \\
  HD 103779 & 296.85 &  -1.02 &  3500 &   62 & $16.17 \pm { 0.03}$ &   \cite{jensen2007variation} &    HD 104705 & 297.45 &  -0.34 &  3500 &   20 & $16.19 \pm { 0.03}$ &   \cite{jensen2007variation} \\
  HD 109399 & 301.71 &  -9.88 &  1900 &  326 & $15.95 \pm { 0.07}$ &   \cite{jensen2007variation} &    HD 122879 & 312.26 &   1.79 &  4800 &  149 & $16.22 \pm { 0.03}$ &   \cite{jensen2007variation} \\
  HD 124314 & 312.67 &  -0.42 &  1100 &    8 & $16.32 \pm { 0.03}$ &   \cite{jensen2007variation} &    HD 147888 & 353.65 &  17.71 &   136 &   41 & $15.83 \pm { 0.04}$ &   \cite{jensen2007variation} \\
  HD 152590 & 344.84 &   1.83 &  1800 &   57 & $16.20 \pm { 0.04}$ &   \cite{jensen2007variation} &    HD 168941 &   5.82 &  -6.31 &  5000 &  549 & $15.87 \pm { 0.13}$ &   \cite{jensen2007variation} \\
  HD 177989 &  17.81 & -11.88 &  5100 & 1049 & $15.83 \pm { 0.04}$ &   \cite{jensen2007variation} &    HD 185418 &  53.60 &  -2.17 &   950 &   35 & $15.94 \pm { 0.04}$ &   \cite{jensen2007variation} \\
  HD 192639 &  74.90 &   1.48 &  1100 &   28 & $16.20 \pm { 0.04}$ &   \cite{jensen2007variation} &    HD 195965 &  85.71 &   5.00 &  1300 &  113 & $15.89 \pm { 0.06}$ &   \cite{jensen2007variation} \\
  HD 202347 &  88.22 &  -2.08 &  1300 &   47 & $15.62 \pm { 0.06}$ &   \cite{jensen2007variation} &    HD 203374 & 100.51 &   8.62 &   820 &  122 & $16.07 \pm { 0.03}$ &   \cite{jensen2007variation} \\
  HD 206267 &  99.29 &   3.74 &  1000 &   65 & $16.05 \pm { 0.06}$ &   \cite{jensen2007variation} &    HD 207198 & 103.14 &   6.99 &  1000 &  121 & $16.08 \pm { 0.03}$ &   \cite{jensen2007variation} \\
  HD 207538 & 101.60 &   4.67 &   880 &   71 & $16.07 \pm { 0.05}$ &   \cite{jensen2007variation} &    HD 209339 & 104.58 &   5.87 &  1100 &  112 & $16.04 \pm { 0.03}$ &   \cite{jensen2007variation} \\
  HD 210839 & 103.83 &   2.61 &   505 &   22 & $16.05 \pm { 0.04}$ &   \cite{jensen2007variation} &    HD 224151 & 115.44 &  -4.64 &  1355 &  109 & $16.30 \pm { 0.04}$ &   \cite{jensen2007variation} \\
  HD 303308 & 287.59 &  -0.61 &  2630 &   27 & $16.34 \pm { 0.06}$ &   \cite{jensen2007variation} &      HD 1383 & 119.02 &  -0.89 &  3340 &   52 & $16.37 \pm { 0.04}$ &   \cite{cartledge2006homogeneity} \\
   HD 12323 & 132.91 &  -5.87 &  2811 &  287 & $16.04 \pm { 0.02}$ &   \cite{cartledge2006homogeneity} &     HD 13268 & 133.96 &  -4.99 &  1693 &  147 & $16.24 \pm { 0.02}$ &   \cite{cartledge2006homogeneity} \\
   HD 14434 & 135.08 &  -3.82 &  2556 &  170 & $16.27 \pm { 0.02}$ &   \cite{cartledge2006homogeneity} &     HD 27778 & 172.76 & -17.39 &   224 &   67 & $15.48 \pm { 0.01}$ &   \cite{cartledge2006homogeneity} \\
   HD 36841 & 204.26 & -17.22 &   451 &  133 & $15.45 \pm { 0.03}$ &   \cite{cartledge2006homogeneity} &     HD 37021 & 209.01 & -19.38 &   402 &  133 & $15.88 \pm { 0.01}$ &   \cite{cartledge2006homogeneity} \\
   HD 37061 & 208.92 & -19.27 &   523 &  172 & $15.79 \pm { 0.02}$ &   \cite{cartledge2006homogeneity} &     HD 37367 & 179.04 &  -1.03 &   988 &   17 & $16.00 \pm { 0.02}$ &   \cite{cartledge2006homogeneity} \\
   HD 37903 & 206.85 & -16.54 &   401 &  114 & $15.63 \pm { 0.03}$ &   \cite{cartledge2006homogeneity} &     HD 43818 & 188.49 &   3.87 &  2570 &  173 & $16.47 \pm { 0.02}$ &   \cite{cartledge2006homogeneity} \\
   HD 52266 & 219.13 &  -0.68 &  1549 &   18 & $16.09 \pm { 0.01}$ &   \cite{cartledge2006homogeneity} &     HD 63005 & 242.47 &  -0.93 & 13661 &  220 & $16.02 \pm { 0.01}$ &   \cite{cartledge2006homogeneity} \\
   HD 71634 & 273.32 & -11.52 &   363 &   72 & $15.79 \pm { 0.02}$ &   \cite{cartledge2006homogeneity} &     HD 75309 & 265.86 &  -1.90 &  2039 &   67 & $15.95 \pm { 0.02}$ &   \cite{cartledge2006homogeneity} \\
   HD 79186 & 267.36 &   2.25 &  1299 &   51 & $16.10 \pm { 0.02}$ &   \cite{cartledge2006homogeneity} &     HD 91824 & 285.70 &   0.07 &  2330 &    2 & $16.13 \pm { 0.01}$ &   \cite{cartledge2006homogeneity} \\
   HD 91983 & 285.88 &   0.05 &  4262 &    3 & $16.19 \pm { 0.03}$ &   \cite{cartledge2006homogeneity} &    HD 111934 & 303.20 &   2.51 &  2251 &   98 & $16.27 \pm { 0.01}$ &   \cite{cartledge2006homogeneity} \\
  HD 116852 & 304.88 & -16.13 & 22727 & 6314 & $15.91 \pm { 0.02}$ &   \cite{cartledge2006homogeneity} &    HD 122879 & 312.26 &   1.79 &  2385 &   74 & $16.23 \pm { 0.01}$ &   \cite{cartledge2006homogeneity} \\
  HD 147888 & 353.65 &  17.71 &    91 &   27 & $15.97 \pm { 0.02}$ &   \cite{cartledge2006homogeneity} &    HD 148594 & 350.93 &  13.94 &   192 &   46 & $15.62 \pm { 0.02}$ &   \cite{cartledge2006homogeneity} \\
  HD 152590 & 344.84 &   1.83 &  1636 &   52 & $16.20 \pm { 0.04}$ &   \cite{cartledge2006homogeneity} &    HD 156110 &  70.99 &  35.71 &   819 &  478 & $15.14 \pm { 0.02}$ &   \cite{cartledge2006homogeneity} \\
  HD 157857 &  12.97 &  13.31 &  3972 &  914 & $16.18 \pm { 0.03}$ &   \cite{cartledge2006homogeneity} &    HD 165955 & 357.41 &  -7.43 &  1205 &  155 & $16.03 \pm { 0.02}$ &   \cite{cartledge2006homogeneity} \\
  HD 175360 &  12.53 & -11.29 &   300 &   58 & $15.60 \pm { 0.02}$ &   \cite{cartledge2006homogeneity} &    HD 185418 &  53.60 &  -2.17 &   754 &   28 & $15.96 \pm { 0.01}$ &   \cite{cartledge2006homogeneity} \\
  HD 190918 &  72.65 &   2.07 &  1954 &   70 & $16.34 \pm { 0.02}$ &   \cite{cartledge2006homogeneity} &    HD 192035 &  83.33 &   7.76 &  2251 &  303 & $15.93 \pm { 0.02}$ &   \cite{cartledge2006homogeneity} \\
  HD 192639 &  74.90 &   1.48 &  2597 &   67 & $16.20 \pm { 0.01}$ &   \cite{cartledge2006homogeneity} &    HD 198478 &  85.75 &   1.49 &  1177 &   30 & $15.95 \pm { 0.02}$ &   \cite{cartledge2006homogeneity} \\
  HD 198781 &  99.94 &  12.61 &   935 &  204 & $15.76 \pm { 0.02}$ &   \cite{cartledge2006homogeneity} &    HD 201345 &  78.44 &  -9.54 &  3189 &  528 & $15.97 \pm { 0.01}$ &   \cite{cartledge2006homogeneity} \\
  HD 203532 & 309.46 & -31.74 &   291 &  153 & $15.58 \pm { 0.01}$ &   \cite{cartledge2006homogeneity} &    HD 206773 &  99.80 &   3.62 &   957 &   60 & $16.00 \pm { 0.01}$ &   \cite{cartledge2006homogeneity} \\
  HD 207198 & 103.14 &   6.99 &  1024 &  124 & $16.08 \pm { 0.01}$ &   \cite{cartledge2006homogeneity} &    HD 208440 & 104.03 &   6.44 &   828 &   92 & $16.05 \pm { 0.02}$ &   \cite{cartledge2006homogeneity} \\
  HD 210809 &  99.85 &  -3.13 &  4323 &  236 & $16.24 \pm { 0.03}$ &   \cite{cartledge2006homogeneity} &    HD 212791 & 101.64 &  -4.30 &   998 &   74 & $15.87 \pm { 0.03}$ &   \cite{cartledge2006homogeneity} \\
  HD 220057 & 112.13 &   0.21 &   391 &    1 & $15.66 \pm { 0.01}$ &   \cite{cartledge2006homogeneity} &    HD 232522 & 130.70 &  -6.71 & 11904 & 1391 & $16.11 \pm { 0.01}$ &   \cite{cartledge2006homogeneity} \\
  HD 308813 & 294.79 &  -1.61 &  5279 &  148 & $16.15 \pm { 0.01}$ &   \cite{cartledge2006homogeneity} &     HD 18100 & 217.90 & -62.70 &  3100 & 2800 & $15.37 \pm { 0.00}$ &   \cite{savage1996interstellargas} \\
  HD 100340 & 258.80 &  61.20 &  5300 & 4600 & $15.69 \pm { 0.00}$ &   \cite{savage1996interstellargas} &     HD 24534 & 163.08 & -17.14 &   810 &  238 & $15.45 \pm { 0.02}$ &   \cite{destree2010detection} \\
  HD 215733 &  85.16 & -36.35 &  3479 & 2062 & $16.10 \pm { 0.05}$ &   \cite{fitzpatrick1997composition} &    HD 167756 & 351.47 & -12.30 &  1977 &  421 & $15.47 \pm { 0.02}$ &   \cite{cardelli1995gas} \\
\bottomrule
\end{tabular}
\label{tab:mg2stars}
\begin{tablenotes}
\item[1)] The distances D are derived from the parallax of stars.
\item[2)] The column densities $\log(N)$ represent the total along the stellar sightlines.
\end{tablenotes}
\end{threeparttable}
\end{table*}

\InterestConflict{The authors declare that they have no conflict of interest.}

\bibliographystyle{scpma}
\bibliography{ms.bib}

@ARTICLE{asplund2009chemical,
       author = {{Asplund}, Martin and {Grevesse}, Nicolas and {Sauval}, A. Jacques and {Scott}, Pat},
        title = "{The Chemical Composition of the Sun}",
      journal = {Annual Review of Astronomy and Astrophysics},
         year = 2009,
        month = sep,
       volume = {47},
       number = {1},
        pages = {481-522},
          doi = {10.1146/annurev.astro.46.060407.145222},
archivePrefix = {arXiv},
       eprint = {0909.0948},
 primaryClass = {astro-ph.SR},
       adsurl = {https://ui.adsabs.harvard.edu/abs/2009ARA&A..47..481A}
}

@article{johnson2019populating,
       author = {{Johnson}, Jennifer A.},
        title = "{Populating the periodic table: Nucleosynthesis of the elements}",
      journal = {Science},
         year = 2019,
        month = feb,
       volume = {363},
       number = {6426},
        pages = {474-478},
          doi = {10.1126/science.aau9540},
       adsurl = {https://ui.adsabs.harvard.edu/abs/2019Sci...363..474J}
}

@ARTICLE{fitzpatrick1997abundance,
       author = {{Fitzpatrick}, Edward L.},
        title = "{The Abundance of Mg in the Interstellar Medium}",
      journal = {Astrophysical Journal, Letters},
         year = 1997,
        month = jun,
       volume = {482},
       number = {2},
        pages = {L199-L202},
          doi = {10.1086/310714},
archivePrefix = {arXiv},
       eprint = {astro-ph/9703162},
 primaryClass = {astro-ph},
       adsurl = {https://ui.adsabs.harvard.edu/abs/1997ApJ...482L.199F}
}

@ARTICLE{jensen2007variation,
       author = {{Jensen}, Adam G. and {Snow}, Theodore P.},
        title = "{The Variation of Magnesium Depletion with Line-of-Sight Conditions}",
      journal = {Astrophysical Journal},
         year = 2007,
        month = nov,
       volume = {669},
       number = {1},
        pages = {401-411},
          doi = {10.1086/521420},
archivePrefix = {arXiv},
       eprint = {0710.1064},
 primaryClass = {astro-ph},
       adsurl = {https://ui.adsabs.harvard.edu/abs/2007ApJ...669..401J}
}

@ARTICLE{gnat2007time,
       author = {{Gnat}, Orly and {Sternberg}, Amiel},
        title = "{Time-dependent Ionization in Radiatively Cooling Gas}",
      journal = {Astrophysical Journal, Supplement},
         year = 2007,
        month = feb,
       volume = {168},
       number = {2},
        pages = {213-230},
          doi = {10.1086/509786},
archivePrefix = {arXiv},
       eprint = {astro-ph/0608181},
 primaryClass = {astro-ph},
       adsurl = {https://ui.adsabs.harvard.edu/abs/2007ApJS..168..213G}
}

@ARTICLE{savage2009extension,
       author = {{Savage}, Blair D. and {Wakker}, Bart P.},
        title = "{The Extension of the Transition Temperature Plasma into the Lower Galactic Halo}",
      journal = {Astrophysical Journal},
         year = 2009,
        month = sep,
       volume = {702},
       number = {2},
        pages = {1472-1489},
          doi = {10.1088/0004-637X/702/2/1472},
archivePrefix = {arXiv},
       eprint = {0907.4955},
 primaryClass = {astro-ph.GA},
       adsurl = {https://ui.adsabs.harvard.edu/abs/2009ApJ...702.1472S}
}

@ARTICLE{wakker2012Characterizing,
       author = {{Wakker}, Bart P. and {Savage}, Blair D. and {Fox}, Andrew J. and
         {Benjamin}, Robert A. and {Shapiro}, Paul R.},
        title = "{Characterizing Transition Temperature Gas in the Galactic Corona}",
      journal = {Astrophysical Journal},
         year = 2012,
        month = apr,
       volume = {749},
       number = {2},
          eid = {157},
        pages = {157},
          doi = {10.1088/0004-637X/749/2/157},
archivePrefix = {arXiv},
       eprint = {1202.5973},
 primaryClass = {astro-ph.GA},
       adsurl = {https://ui.adsabs.harvard.edu/abs/2012ApJ...749..157W}
}

@article{bergeron1994hubble,
       author = {{Bergeron}, Jacqueline and {Petitjean}, Patrick and {Sargent}, W.~L.~W. and {Bahcall}, John N. and {Boksenberg}, Alec and {Hartig}, George F. and {Jannuzi}, Buell T. and {Kirhakos}, Sofia and {Savage}, Blair D. and {Schneider}, Donald P. and {Turnshek}, David A. and {Weymann}, Ray J. and {Wolfe}, Arthur M.},
        title = "{The Hubble Space Telescope Quasar Absorption Line Key Project. VI. Properties of the Metal-rich Systems}",
      journal = {Astrophysical Journal},
         year = 1994,
        month = nov,
       volume = {436},
        pages = {33},
          doi = {10.1086/174878},
       adsurl = {https://ui.adsabs.harvard.edu/abs/1994ApJ...436...33B}
}

@ARTICLE{wolfire1995neutral,
  title={The neutral atomic phases of the interstellar medium},
  author={Wolfire, Mark G and Hollenbach, David and McKee, Christopher F and Tielens, AGGM and Bakes, ELO},
  journal={Astrophysical Journal},
  volume={443},
  pages={152--168},
  year={1995},
  doi={10.1086/175122}
}

@ARTICLE{ferriere2001interstellar,
  title={The interstellar environment of our galaxy},
  author={Ferriere, Katia M},
  journal={Reviews of Modern Physics},
  volume={73},
  number={4},
  pages={1031},
  year={2001},
  doi={10.1103/RevModPhys.73.1031},
  publisher={APS}
}

@article{chen2017mg,
       author = {{Chen}, Shi-Fan S. and {Simcoe}, Robert A. and {Torrey}, Paul and {Ba{\~n}ados}, Eduardo and {Cooksey}, Kathy and {Cooper}, Tom and {Furesz}, Gabor and {Matejek}, Michael and {Miller}, Daniel and {Turner}, Monica and {Venemans}, Bram and {Decarli}, Roberto and {Farina}, Emanuele P. and {Mazzucchelli}, Chiara and {Walter}, Fabian},
        title = "{Mg II Absorption at 2 < Z < 7 with Magellan/Fire. III. Full Statistics of Absorption toward 100 High-redshift QSOs}",
      journal = {Astrophysical Journal},
         year = 2017,
        month = dec,
       volume = {850},
       number = {2},
          eid = {188},
        pages = {188},
          doi = {10.3847/1538-4357/aa9707},
archivePrefix = {arXiv},
       eprint = {1612.02829},
 primaryClass = {astro-ph.CO},
       adsurl = {https://ui.adsabs.harvard.edu/abs/2017ApJ...850..188C}
}

@ARTICLE{kacprzak2008halo,
       author = {{Kacprzak}, Glenn G. and {Churchill}, Christopher W. and {Steidel}, Charles C. and {Murphy}, Michael T.},
        title = "{Halo Gas Cross Sections and Covering Fractions of Mg II Absorption Selected Galaxies}",
      journal = {Astronomical Journal},
         year = 2008,
        month = mar,
       volume = {135},
       number = {3},
        pages = {922-927},
          doi = {10.1088/0004-6256/135/3/922},
archivePrefix = {arXiv},
       eprint = {0710.5765},
 primaryClass = {astro-ph},
       adsurl = {https://ui.adsabs.harvard.edu/abs/2008AJ....135..922K}
}

@ARTICLE{chen2010empirical,
       author = {{Chen}, Hsiao-Wen and {Helsby}, Jennifer E. and {Gauthier}, Jean-Ren{\'e} and {Shectman}, Stephen A. and {Thompson}, Ian B. and {Tinker}, Jeremy L.},
        title = "{An Empirical Characterization of Extended Cool Gas Around Galaxies Using Mg II Absorption Features}",
      journal = {Astrophysical Journal},
         year = 2010,
        month = may,
       volume = {714},
       number = {2},
        pages = {1521-1541},
          doi = {10.1088/0004-637X/714/2/1521},
archivePrefix = {arXiv},
       eprint = {1004.0705},
 primaryClass = {astro-ph.CO},
       adsurl = {https://ui.adsabs.harvard.edu/abs/2010ApJ...714.1521C}
}

@article{nielsen2013magiicat,
       author = {{Nielsen}, Nikole M. and {Churchill}, Christopher W. and {Kacprzak}, Glenn G. and {Murphy}, Michael T.},
        title = "{MAGIICAT I. The Mg II Absorber-Galaxy Catalog}",
      journal = {Astrophysical Journal},
         year = 2013,
        month = oct,
       volume = {776},
       number = {2},
          eid = {114},
        pages = {114},
          doi = {10.1088/0004-637X/776/2/114},
archivePrefix = {arXiv},
       eprint = {1304.6716},
 primaryClass = {astro-ph.CO},
       adsurl = {https://ui.adsabs.harvard.edu/abs/2013ApJ...776..114N}
}

@ARTICLE{Lanzetta1990intermediate,
       author = {{Lanzetta}, Kenneth M. and {Bowen}, David},
        title = "{Intermediate-Redshift Galaxy Halos: Results from QSO Absorption Lines}",
      journal = {Astrophysical Journal},
         year = 1990,
        month = jul,
       volume = {357},
        pages = {321},
          doi = {10.1086/168922},
       adsurl = {https://ui.adsabs.harvard.edu/abs/1990ApJ...357..321L}
}

@ARTICLE{bergeron1991sample,
       author = {{Bergeron}, J. and {Boiss{\'e}}, P.},
        title = "{A sample of galaxies giving rise to Mg II quasar absorption systems.}",
      journal = {Astronomy and Astrophysics},
         year = 1991,
        month = mar,
       volume = {243},
        pages = {344},
       adsurl = {https://ui.adsabs.harvard.edu/abs/1991A&A...243..344B}
}

@INPROCEEDINGS{steidel1995nature,
       author = {{Steidel}, C.~C.},
        title = "{The Nature and Evolution of Absorption-Selected Galaxies}",
    booktitle = {QSO Absorption Lines},
         year = 1995,
       editor = {{Meylan}, Georges},
        month = jan,
        pages = {139},
archivePrefix = {arXiv},
       eprint = {astro-ph/9509098},
 primaryClass = {astro-ph},
       adsurl = {https://ui.adsabs.harvard.edu/abs/1995qal..conf..139S}
}

@ARTICLE{bouche2006new,
       author = {{Bouch{\'e}}, Nicolas and {Murphy}, Michael T. and {P{\'e}roux}, C{\'e}line and {Csabai}, Istv{\'a}n and {Wild}, Vivienne},
        title = "{New perspectives on strong z≃ 0.5 Mg ii absorbers: are halo mass and equivalent width anticorrelated?}",
      journal = {Monthly Notices of the Royal Astronomical Society},
         year = 2006,
        month = sep,
       volume = {371},
       number = {1},
        pages = {495-512},
          doi = {10.1111/j.1365-2966.2006.10685.x},
archivePrefix = {arXiv},
       eprint = {astro-ph/0606328},
 primaryClass = {astro-ph},
       adsurl = {https://ui.adsabs.harvard.edu/abs/2006MNRAS.371..495B}
}

@article{lan2020coevolution,
       author = {{Lan}, Ting-Wen},
        title = "{The Coevolution of Galaxies and the Cool Circumgalactic Medium Probed with the SDSS and DESI Legacy Imaging Surveys}",
      journal = {Astrophysical Journal},
         year = 2020,
        month = jul,
       volume = {897},
       number = {1},
          eid = {97},
        pages = {97},
          doi = {10.3847/1538-4357/ab989a},
archivePrefix = {arXiv},
       eprint = {1911.01271},
 primaryClass = {astro-ph.GA},
       adsurl = {https://ui.adsabs.harvard.edu/abs/2020ApJ...897...97L}
}

@ARTICLE{hopkins2006evolution,
       author = {{Hopkins}, Philip F. and {Hernquist}, Lars and {Cox}, Thomas J. and {Robertson}, Brant and {Di Matteo}, Tiziana and {Springel}, Volker},
        title = "{The Evolution in the Faint-End Slope of the Quasar Luminosity Function}",
      journal = {Astrophysical Journal},
         year = 2006,
        month = mar,
       volume = {639},
       number = {2},
        pages = {700-709},
          doi = {10.1086/499351},
archivePrefix = {arXiv},
       eprint = {astro-ph/0508299},
 primaryClass = {astro-ph},
       adsurl = {https://ui.adsabs.harvard.edu/abs/2006ApJ...639..700H}
}

@ARTICLE{york2000sdss,
       author = {{York}, Donald G. and {Adelman}, J. and {Anderson}, Jr., John E. and others},
        title = "{The Sloan Digital Sky Survey: Technical Summary}",
      journal = {Astronomical Journal},
         year = 2000,
        month = sep,
       volume = {120},
       number = {3},
        pages = {1579-1587},
          doi = {10.1086/301513},
archivePrefix = {arXiv},
       eprint = {astro-ph/0006396},
 primaryClass = {astro-ph},
       adsurl = {https://ui.adsabs.harvard.edu/abs/2000AJ....120.1579Y}
}

@ARTICLE{desi2016desi,
       author = {{DESI Collaboration} and others},
        title = "{The DESI Experiment Part II: Instrument Design}",
      journal = {arXiv e-prints},
         year = 2016,
        month = oct,
          eid = {arXiv:1611.00037},
        pages = {arXiv:1611.00037},
          doi = {10.48550/arXiv.1611.00037},
archivePrefix = {arXiv},
       eprint = {1611.00037},
 primaryClass = {astro-ph.IM},
       adsurl = {https://ui.adsabs.harvard.edu/abs/2016arXiv161100037D}
}

@article{herenz2013milky,
       author = {{Herenz}, P. and {Richter}, P. and {Charlton}, J.~C. and {Masiero}, J.~R.},
        title = "{The Milky Way halo as a QSO absorption-line system. New results from an HST/STIS absorption-line catalogue of Galactic high-velocity clouds}",
      journal = {Astronomy and Astrophysics},
         year = 2013,
        month = feb,
       volume = {550},
          eid = {A87},
        pages = {A87},
          doi = {10.1051/0004-6361/201220531},
archivePrefix = {arXiv},
       eprint = {1301.1345},
 primaryClass = {astro-ph.CO},
       adsurl = {https://ui.adsabs.harvard.edu/abs/2013A&A...550A..87H}
}

@ARTICLE{morton2003atomic,
       author = {{Morton}, Donald C.},
        title = "{Atomic Data for Resonance Absorption Lines. III. Wavelengths Longward of the Lyman Limit for the Elements Hydrogen to Gallium}",
      journal = {Astrophysical Journal, Supplement},
         year = 2003,
        month = nov,
       volume = {149},
       number = {1},
        pages = {205-238},
          doi = {10.1086/377639},
       adsurl = {https://ui.adsabs.harvard.edu/abs/2003ApJS..149..205M}
}

@ARTICLE{savage1996interstellargas,
       author = {{Savage}, Blair D. and {Sembach}, Kenneth R.},
        title = "{Interstellar Gas-Phase Abundances and Physical Conditions toward Two Distant High-Latitude Halo Stars}",
      journal = {Astrophysical Journal},
         year = 1996,
        month = oct,
       volume = {470},
        pages = {893},
          doi = {10.1086/177919},
       adsurl = {https://ui.adsabs.harvard.edu/abs/1996ApJ...470..893S}
}

@ARTICLE{destree2010detection,
       author = {{Destree}, Joshua D. and {Williamson}, Karen E. and {Snow}, Theodore P.},
        title = "{Detection in the Interstellar Medium of the Weak [Mg II] Transition at 1398.8 {\r{A}}}",
      journal = {Astrophysical Journal},
         year = 2010,
        month = mar,
       volume = {712},
       number = {1},
        pages = {L48-L52},
          doi = {10.1088/2041-8205/712/1/L48},
       adsurl = {https://ui.adsabs.harvard.edu/abs/2010ApJ...712L..48D}
}

@ARTICLE{cardelli1995gas,
       author = {{Cardelli}, Jason A. and {Sembach}, Kenneth R. and {Savage}, Blair D.},
        title = "{Gas Phase Abundances and Conditions along the Sight Line to the Low-Halo, Inner Galaxy Star HD 167756}",
      journal = {Astrophysical Journal},
         year = 1995,
        month = feb,
       volume = {440},
        pages = {241},
          doi = {10.1086/175265},
       adsurl = {https://ui.adsabs.harvard.edu/abs/1995ApJ...440..241C}
}

@MISC{peeples2017hubble,
       author = {{Peeples}, M. and {Tumlinson}, J. and {Fox}, A. and {Aloisi}, A. and {Fleming}, S. and {Jedrzejewski}, R. and {Oliveira}, C. and {Ayres}, T. and {Danforth}, C. and {Keeney}, B. and {Jenkins}, E.},
        title = "{The Hubble Spectroscopic Legacy Archive}",
 howpublished = {Instrument Science Report COS 2017-4},
         year = 2017,
        month = apr,
        pages = {4},
       adsurl = {https://ui.adsabs.harvard.edu/abs/2017cos..rept....4P}
}

@MISC{ghavamian2009cos,
       author = {{Ghavamian}, Parviz},
        title = "{COS FUV Line Spread Function Characterization}",
 howpublished = {HST Proposal},
         year = 2009,
        month = jul,
        pages = {12010},
       adsurl = {https://ui.adsabs.harvard.edu/abs/2009hst..prop12010G}
}

@BOOK{mihalas1981galactic,
       author = {{Mihalas}, D. and {Binney}, J.},
        title = "{Galactic astronomy. Structure and kinematics}",
         year = 1981,
       adsurl = {https://ui.adsabs.harvard.edu/abs/1981gask.book.....M}
}

@ARTICLE{Roamn-Duval2021Metal,
       author = {{Roman-Duval}, Julia and {Jenkins}, Edward B. and {Tchernyshyov}, Kirill and {Williams}, Benjamin and {Clark}, Christopher J.~R. and {Gordon}, Karl D. and {Meixner}, Margaret and {Hagen}, Lea and {Peek}, Joshua and {Sandstrom}, Karin and {Werk}, Jessica and {Yanchulova Merica-Jones}, Petia},
        title = "{METAL: The Metal Evolution, Transport, and Abundance in the Large Magellanic Cloud Hubble Program. II. Variations of Interstellar Depletions and Dust-to-gas Ratio within the LMC}",
      journal = {Astrophysical Journal},
         year = 2021,
        month = apr,
       volume = {910},
       number = {2},
          eid = {95},
        pages = {95},
          doi = {10.3847/1538-4357/abdeb6},
archivePrefix = {arXiv},
       eprint = {2101.09399},
 primaryClass = {astro-ph.GA},
       adsurl = {https://ui.adsabs.harvard.edu/abs/2021ApJ...910...95R}
}

@INBOOK{soderblom2021cos,
       author = {{Soderblom}, David R.},
        title = "{COS Data Handbook v. 5.0}",
    booktitle = {COS Data Handbook},
         year = 2021,
        pages = {5},
       adsurl = {https://ui.adsabs.harvard.edu/abs/2021cosd.book....5S}
}

@ARTICLE{churchill2020mgii,
       author = {{Churchill}, Christopher W. and {Evans}, Jessica L. and {Stemock}, Bryson and {Nielsen}, Nikole M. and {Kacprzak}, Glenn G. and {Murphy}, Michael T.},
        title = "{Mg II Absorbers in High-resolution Quasar Spectra. I. Voigt Profile Models}",
      journal = {Astrophysical Journal},
         year = 2020,
        month = nov,
       volume = {904},
       number = {1},
          eid = {28},
        pages = {28},
          doi = {10.3847/1538-4357/abbb34},
archivePrefix = {arXiv},
       eprint = {2008.08487},
 primaryClass = {astro-ph.GA},
       adsurl = {https://ui.adsabs.harvard.edu/abs/2020ApJ...904...28C}
}

@ARTICLE{lenz2017new,
       author = {{Lenz}, Daniel and {Hensley}, Brandon S. and {Dor{\'e}}, Olivier},
        title = "{A New, Large-scale Map of Interstellar Reddening Derived from H I Emission}",
      journal = {Astrophysical Journal},
         year = 2017,
        month = sep,
       volume = {846},
       number = {1},
          eid = {38},
        pages = {38},
          doi = {10.3847/1538-4357/aa84af},
archivePrefix = {arXiv},
       eprint = {1706.00011},
 primaryClass = {astro-ph.GA},
       adsurl = {https://ui.adsabs.harvard.edu/abs/2017ApJ...846...38L}
}

@ARTICLE{Murray1984Interstellar,
       author = {{Murray}, M.~J. and {Dufton}, P.~L. and {Hibbert}, A. and {York}, D.~G.},
        title = "{Interstellar magnesium abundances.}",
      journal = {Astrophysical Journal},
         year = 1984,
        month = jul,
       volume = {282},
        pages = {481-484},
          doi = {10.1086/162225},
       adsurl = {https://ui.adsabs.harvard.edu/abs/1984ApJ...282..481M}
}

@ARTICLE{Jenkins1986Abundances,
       author = {{Jenkins}, E.~B. and {Savage}, B.~D. and {Spitzer}, L., Jr.},
        title = "{Abundances of Interstellar Atoms from Ultraviolet Absorption Lines}",
      journal = {Astrophysical Journal},
         year = 1986,
        month = feb,
       volume = {301},
        pages = {355},
          doi = {10.1086/163906},
       adsurl = {https://ui.adsabs.harvard.edu/abs/1986ApJ...301..355J}
}

@article{bekhti2016hi4pi,
author = {{HI4PI Collaboration} and {Ben Bekhti}, N. and {Fl{\"o}er}, L. and {Keller}, R. and {Kerp}, J. and {Lenz}, D. and {Winkel}, B. and {Bailin}, J. and {Calabretta}, M.~R. and {Dedes}, L. and {Ford}, H.~A. and {Gibson}, B.~K. and {Haud}, U. and {Janowiecki}, S. and {Kalberla}, P.~M.~W. and {Lockman}, F.~J. and {McClure-Griffiths}, N.~M. and {Murphy}, T. and {Nakanishi}, H. and {Pisano}, D.~J. and {Staveley-Smith}, L.},
        title = "{HI4PI: A full-sky H I survey based on EBHIS and GASS}",
      journal = {Astronomy and Astrophysics},
         year = 2016,
        month = oct,
       volume = {594},
          eid = {A116},
        pages = {A116},
          doi = {10.1051/0004-6361/201629178},
archivePrefix = {arXiv},
       eprint = {1610.06175},
 primaryClass = {astro-ph.GA},
       adsurl = {https://ui.adsabs.harvard.edu/abs/2016A&A...594A.116H}
}

@ARTICLE{Kalberla2020HI,
       author = {{Kalberla}, P.~M.~W. and {Kerp}, J. and {Haud}, U.},
        title = "{H I filaments are cold and associated with dark molecular gas. HI4PI-based estimates of the local diffuse CO-dark H$_{2}$ distribution}",
      journal = {Astronomy and Astrophysics},
         year = 2020,
        month = jul,
       volume = {639},
          eid = {A26},
        pages = {A26},
          doi = {10.1051/0004-6361/202037602},
archivePrefix = {arXiv},
       eprint = {2004.14630},
 primaryClass = {astro-ph.GA},
       adsurl = {https://ui.adsabs.harvard.edu/abs/2020A&A...639A..26K}
}

@article{wakker2011measuring,
  title={MEASURING TURBULENCE IN THE INTERSTELLAR MEDIUM BY COMPARING N (H i; Ly$\alpha$) AND N (H i; 21 cm)},
  author={Wakker, Bart P and Lockman, Felix J and Brown, Jonathan M},
  journal={Astrophysical Journal},
  volume={728},
  number={2},
  pages={159},
  year={2011},
  doi={10.1088/0004-637X/728/2/159},
  publisher={IOP Publishing}
}

@article{nielsen2013magiicatII,
      author = {{Nielsen}, Nikole M. and {Churchill}, Christopher W. and {Kacprzak}, Glenn G.},
        title = "{MAGIICAT II. General Characteristics of the Mg II Absorbing Circumgalactic Medium}",
      journal = {Astrophysical Journal},
         year = 2013,
        month = oct,
       volume = {776},
       number = {2},
          eid = {115},
        pages = {115},
          doi = {10.1088/0004-637X/776/2/115},
archivePrefix = {arXiv},
       eprint = {1211.1380},
 primaryClass = {astro-ph.CO},
       adsurl = {https://ui.adsabs.harvard.edu/abs/2013ApJ...776..115N}
}

@article{werk2013cos,
       author = {{Werk}, Jessica K. and {Prochaska}, J. Xavier and {Thom}, Christopher and {Tumlinson}, Jason and {Tripp}, Todd M. and {O'Meara}, John M. and {Peeples}, Molly S.},
        title = "{The COS-Halos Survey: An Empirical Description of Metal-line Absorption in the Low-redshift Circumgalactic Medium}",
      journal = {Astrophysical Journal, Supplement},
         year = 2013,
        month = feb,
       volume = {204},
       number = {2},
          eid = {17},
        pages = {17},
          doi = {10.1088/0067-0049/204/2/17},
archivePrefix = {arXiv},
       eprint = {1212.0558},
 primaryClass = {astro-ph.CO},
       adsurl = {https://ui.adsabs.harvard.edu/abs/2013ApJS..204...17W}
}

@article{lehner2015evidence,
       author = {{Lehner}, Nicolas and {Howk}, J. Christopher and {Wakker}, Bart P.},
        title = "{Evidence for a Massive, Extended Circumgalactic Medium Around the Andromeda Galaxy}",
      journal = {Astrophysical Journal},
         year = 2015,
        month = may,
       volume = {804},
       number = {2},
          eid = {79},
        pages = {79},
          doi = {10.1088/0004-637X/804/2/79},
archivePrefix = {arXiv},
       eprint = {1404.6540},
 primaryClass = {astro-ph.GA},
       adsurl = {https://ui.adsabs.harvard.edu/abs/2015ApJ...804...79L}
}

@article{keeney2017characterizing,
       author = {{Keeney}, Brian A. and {Stocke}, John T. and {Danforth}, Charles W. and {Shull}, J. Michael and {Pratt}, Cameron T. and {Froning}, Cynthia S. and {Green}, James C. and {Penton}, Steven V. and {Savage}, Blair D.},
        title = "{Characterizing the Circumgalactic Medium of Nearby Galaxies with HST/COS and HST/STIS Absorption-line Spectroscopy. II. Methods and Models}",
      journal = {Astrophysical Journal, Supplement},
         year = 2017,
        month = may,
       volume = {230},
       number = {1},
          eid = {6},
        pages = {6},
          doi = {10.3847/1538-4365/aa6b59},
archivePrefix = {arXiv},
       eprint = {1704.00235},
 primaryClass = {astro-ph.GA},
       adsurl = {https://ui.adsabs.harvard.edu/abs/2017ApJS..230....6K}
}

@article{keeney2012significance,
       author = {{Keeney}, Brian A. and {Danforth}, Charles W. and {Stocke}, John T. and {France}, Kevin and {Green}, James C.},
        title = "{On the Significance of Absorption Features in HST/COS Data}",
      journal = {Publications of the Astronomical Society of the Pacific},
         year = 2012,
        month = aug,
       volume = {124},
       number = {918},
        pages = {830},
          doi = {10.1086/667392},
archivePrefix = {arXiv},
       eprint = {1206.2951},
 primaryClass = {astro-ph.IM},
       adsurl = {https://ui.adsabs.harvard.edu/abs/2012PASP..124..830K}
}

@article{cameron2011estimation,
  title={On the estimation of confidence intervals for binomial population proportions in astronomy: the simplicity and superiority of the Bayesian approach},
  author={Cameron, Ewan},
  journal={Publications of the Astronomical Society of Australia},
  volume={28},
  number={2},
  pages={128--139},
  year={2011},
  doi={10.1071/AS10012},
  publisher={Cambridge University Press}
}

@ARTICLE{bish2019galactic,
       author = {{Bish}, Hannah V. and {Werk}, Jessica K. and {Prochaska}, J. Xavier and
         {Rubin}, Kate H.~R. and {Zheng}, Yong and {O'Meara}, John M. and
         {Deason}, Alis J.},
        title = "{Galactic Gas Flows from Halo to Disk: Tomography and Kinematics at the Milky Way{\textquoteright}s Disk-Halo Interface}",
      journal = {Astrophysical Journal},
         year = 2019,
        month = sep,
       volume = {882},
       number = {2},
          eid = {76},
        pages = {76},
          doi = {10.3847/1538-4357/ab3414},
archivePrefix = {arXiv},
       eprint = {1907.09459},
 primaryClass = {astro-ph.GA},
       adsurl = {https://ui.adsabs.harvard.edu/abs/2019ApJ...882...76B}
}

@ARTICLE{churchill2003physical,
       author = {{Churchill}, Christopher W. and {Vogt}, Steven S. and {Charlton}, Jane C.},
        title = "{The Physical Conditions of Intermediate-Redshift Mg II Absorbing Clouds from Voigt Profile Analysis}",
      journal = {Astronomical Journal},
         year = 2003,
        month = jan,
       volume = {125},
       number = {1},
        pages = {98-115},
          doi = {10.1086/345513},
archivePrefix = {arXiv},
       eprint = {astro-ph/0210196},
 primaryClass = {astro-ph},
       adsurl = {https://ui.adsabs.harvard.edu/abs/2003AJ....125...98C}
}

@article{vazquez1994hierarchical,
  title={Hierarchical structure in nearly pressureless flows as a consequence of self-similar statistics},
  author={Vazquez-Semadeni, Enrique},
  journal={Astrophysical Journal},
  volume={423},
  pages={681},
  year={1994},
  doi={10.1086/173707}
}

@ARTICLE{ostriker2001density,
       author = {{Ostriker}, Eve C. and {Stone}, James M. and {Gammie}, Charles F.},
        title = "{Density, Velocity, and Magnetic Field Structure in Turbulent Molecular Cloud Models}",
      journal = {Astrophysical Journal},
         year = 2001,
        month = jan,
       volume = {546},
       number = {2},
        pages = {980-1005},
          doi = {10.1086/318290},
archivePrefix = {arXiv},
       eprint = {astro-ph/0008454},
 primaryClass = {astro-ph},
       adsurl = {https://ui.adsabs.harvard.edu/abs/2001ApJ...546..980O}
}

@ARTICLE{fitzpatrick1997composition,
       author = {{Fitzpatrick}, Edward L. and {Spitzer}, Lyman, Jr.},
        title = "{Composition of Interstellar Clouds in the Disk and Halo. IV. HD 215733}",
      journal = {Astrophysical Journal},
         year = 1997,
        month = feb,
       volume = {475},
       number = {2},
        pages = {623-641},
          doi = {10.1086/303556},
       adsurl = {https://ui.adsabs.harvard.edu/abs/1997ApJ...475..623F}
}

@ARTICLE{cartledge2006homogeneity,
       author = {{Cartledge}, Stefan I.~B. and {Lauroesch}, J.~T. and {Meyer}, David M. and {Sofia}, Ulysses J.},
        title = "{The Homogeneity of Interstellar Elemental Abundances in the Galactic Disk}",
      journal = {Astrophysical Journal},
         year = 2006,
        month = apr,
       volume = {641},
       number = {1},
        pages = {327-346},
          doi = {10.1086/500297},
archivePrefix = {arXiv},
       eprint = {astro-ph/0512312},
 primaryClass = {astro-ph},
       adsurl = {https://ui.adsabs.harvard.edu/abs/2006ApJ...641..327C}
}

@ARTICLE{Foreman2013emcee,
       author = {{Foreman-Mackey}, Daniel and {Hogg}, David W. and {Lang}, Dustin and {Goodman}, Jonathan},
        title = "{emcee: The MCMC Hammer}",
      journal = {Publications of the Astronomical Society of the Pacific},
         year = 2013,
        month = mar,
       volume = {125},
       number = {925},
        pages = {306},
          doi = {10.1086/670067},
archivePrefix = {arXiv},
       eprint = {1202.3665},
 primaryClass = {astro-ph.IM},
       adsurl = {https://ui.adsabs.harvard.edu/abs/2013PASP..125..306F}
}

@article{Cherrey2023MusEGF,
  title={MusE GAs FLOw and Wind (MEGAFLOW) X. The cool gas and covering fraction of Mg  ii in galaxy groups},
  author={Maxime Cherrey and Nicolas F. Bouch'e and Johannes Zabl and Ilane Schroetter and Martin Wendt and Ivanna Langan and Johan Richard and Joop Schaye and Wilfried Mercier and Beno{\^i}t {\'E}pinat and Thierry Contini},
  journal={Monthly Notices of the Royal Astronomical Society},
  year={2023},
  url={https://api.semanticscholar.org/CorpusID:265608802}
}

@ARTICLE{bennett2019vertical,
       author = {{Bennett}, Morgan and {Bovy}, Jo},
        title = "{Vertical waves in the solar neighbourhood in Gaia DR2}",
      journal = {Monthly Notices of the Royal Astronomical Society},
         year = 2019,
        month = jan,
       volume = {482},
       number = {1},
        pages = {1417-1425},
          doi = {10.1093/mnras/sty2813},
archivePrefix = {arXiv},
       eprint = {1809.03507},
 primaryClass = {astro-ph.GA},
       adsurl = {https://ui.adsabs.harvard.edu/abs/2019MNRAS.482.1417B}
}

@ARTICLE{zheng2019revealing,
       author = {{Zheng}, Y. and {Peek}, J.~E.~G. and {Putman}, M.~E. and {Werk}, J.~K.},
        title = "{Revealing the Milky Way{\textquoteright}s Hidden Circumgalactic Medium with the Cosmic Origins Spectrograph Quasar Database for Galactic Absorption Lines}",
      journal = {Astrophysical Journal},
         year = 2019,
        month = jan,
       volume = {871},
       number = {1},
          eid = {35},
        pages = {35},
          doi = {10.3847/1538-4357/aaf6eb},
archivePrefix = {arXiv},
       eprint = {1710.10703},
 primaryClass = {astro-ph.GA},
       adsurl = {https://ui.adsabs.harvard.edu/abs/2019ApJ...871...35Z}
}

@ARTICLE{qu2019warm,
       author = {{Qu}, Zhijie and {Bregman}, Joel N.},
        title = "{The Warm Gaseous Disk and the Anisotropic Circumgalactic Medium of the Milky Way}",
      journal = {Astrophysical Journal},
         year = 2019,
        month = aug,
       volume = {880},
       number = {2},
          eid = {89},
        pages = {89},
          doi = {10.3847/1538-4357/ab2a0b},
archivePrefix = {arXiv},
       eprint = {1906.12259},
 primaryClass = {astro-ph.GA},
       adsurl = {https://ui.adsabs.harvard.edu/abs/2019ApJ...880...89Q}
}

@ARTICLE{snowden1997rosat,
       author = {{Snowden}, S.~L. and {Egger}, R. and {Freyberg}, M.~J. and {McCammon}, D. and {Plucinsky}, P.~P. and {Sanders}, W.~T. and {Schmitt}, J.~H.~M.~M. and {Tr{\"u}mper}, J. and {Voges}, W.},
        title = "{ROSAT Survey Diffuse X-Ray Background Maps. II.}",
      journal = {Astrophysical Journal},
         year = 1997,
        month = aug,
       volume = {485},
       number = {1},
        pages = {125-135},
          doi = {10.1086/304399},
       adsurl = {https://ui.adsabs.harvard.edu/abs/1997ApJ...485..125S}
}

@ARTICLE{savage2003distribution,
       author = {{Savage}, B.~D. and {Sembach}, K.~R. and {Wakker}, B.~P. and {Richter}, P. and {Meade}, M. and {Jenkins}, E.~B. and {Shull}, J.~M. and {Moos}, H.~W. and {Sonneborn}, G.},
        title = "{Distribution and Kinematics of O VI in the Galactic Halo}",
      journal = {Astrophysical Journal, Supplement},
         year = 2003,
        month = may,
       volume = {146},
       number = {1},
        pages = {125-164},
          doi = {10.1086/346229},
archivePrefix = {arXiv},
       eprint = {astro-ph/0208140},
 primaryClass = {astro-ph},
       adsurl = {https://ui.adsabs.harvard.edu/abs/2003ApJS..146..125S}
}

@ARTICLE{an2019asymmetric,
       author = {{An}, Deokkeun},
        title = "{Asymmetric Mean Metallicity Distribution of the Milky Way{\textquoteright}s Disk}",
      journal = {Astrophysical Journal},
         year = 2019,
        month = jun,
       volume = {878},
       number = {2},
          eid = {L31},
        pages = {L31},
          doi = {10.3847/2041-8213/ab2467},
archivePrefix = {arXiv},
       eprint = {1906.01244},
 primaryClass = {astro-ph.GA},
       adsurl = {https://ui.adsabs.harvard.edu/abs/2019ApJ...878L..31A}
}

@ARTICLE{thomas2024spectroTranslator,
	author = {{Thomas}, G.~F. and {Battaglia}, G. and {Gran}, F. and {Fern{\'a}ndez-Alvar}, E. and {Tsantaki}, M. and {Pancino}, E. and {Hill}, V. and {Kordopatis}, G. and {Gallart}, C. and {Turchi}, A. and {Masseron}, T.},
	title = "{SpectroTranslator: Deep-neural network algorithm for homogenising spectroscopic parameters}",
	journal = {Astronomy and Astrophysics},
	year = 2024,
	month = oct,
	volume = {690},
	eid = {A54},
	pages = {A54},
	doi = {10.1051/0004-6361/202450198},
	archivePrefix = {arXiv},
	eprint = {2404.02578},
	primaryClass = {astro-ph.GA},
	adsurl = {https://ui.adsabs.harvard.edu/abs/2024A&A...690A..54T}
}

@article{tumlinson2017circumgalactic,
       author = {{Tumlinson}, Jason and {Peeples}, Molly S. and {Werk}, Jessica K.},
        title = "{The Circumgalactic Medium}",
      journal = {Annual Review of Astronomy and Astrophysics},
         year = 2017,
        month = aug,
       volume = {55},
       number = {1},
        pages = {389-432},
          doi = {10.1146/annurev-astro-091916-055240},
archivePrefix = {arXiv},
       eprint = {1709.09180},
 primaryClass = {astro-ph.GA},
       adsurl = {https://ui.adsabs.harvard.edu/abs/2017ARA&A..55..389T}
}

@article{lockman1991vertical,
  title={Vertical distribution and support of galactic HI},
  author={Lockman, Felix J and Gehman, Curtis S},
  journal={Astrophysical Journal},
  volume={382},
  pages={182--188},
  year={1991},
  doi={10.1086/170750}
}

@INBOOK{wakker2013high,
       author = {{Wakker}, Bart P. and {van Woerden}, Hugo},
        title = "{High-Velocity Clouds}",
    booktitle = {Planets, Stars and Stellar Systems. Volume 5: Galactic Structure and Stellar Populations},
         year = 2013,
       editor = {{Oswalt}, Terry D. and {Gilmore}, Gerard},
       volume = {5},
        pages = {587},
          doi = {10.1007/978-94-007-5612-0\_12},
       adsurl = {https://ui.adsabs.harvard.edu/abs/2013pss5.book..587W}
}

@ARTICLE{steidel1994field,
       author = {{Steidel}, Charles C. and {Dickinson}, Mark and {Persson}, S.~E.},
        title = "{Field Galaxy Evolution Since Z approximately 1 from a Sample of QSO Absorption--selected Galaxies}",
      journal = {Astrophysical Journal},
         year = 1994,
        month = dec,
       volume = {437},
        pages = {L75},
          doi = {10.1086/187686},
archivePrefix = {arXiv},
       eprint = {astro-ph/9410025},
 primaryClass = {astro-ph},
       adsurl = {https://ui.adsabs.harvard.edu/abs/1994ApJ...437L..75S}
}

@INPROCEEDINGS{churchill2005mgii,
       author = {{Churchill}, Christopher W. and {Kacprzak}, Glenn G. and {Steidel}, Charles C.},
        title = "{MgII absorption through intermediate redshift galaxies}",
    booktitle = {IAU Colloq. 199: Probing Galaxies through Quasar Absorption Lines},
         year = 2005,
       editor = {{Williams}, Peter and {Shu}, Cheng-Gang and {Menard}, Brice},
        month = mar,
        pages = {24-41},
          doi = {10.1017/S1743921305002401},
archivePrefix = {arXiv},
       eprint = {astro-ph/0504392},
 primaryClass = {astro-ph},
       adsurl = {https://ui.adsabs.harvard.edu/abs/2005pgqa.conf...24C}
}

@ARTICLE{savage1997absorption,
       author = {{Savage}, Blair D. and {Sembach}, Kenneth R. and {Lu}, Limin},
        title = "{Absorption by Highly Ionized Interstellar Gas Along Extragalactic and Galactic Sight Lines}",
      journal = {Astronomical Journal},
         year = 1997,
        month = jun,
       volume = {113},
        pages = {2158},
          doi = {10.1086/118427},
       adsurl = {https://ui.adsabs.harvard.edu/abs/1997AJ....113.2158S}
}

@article{dickey1990hi,
  title={HI in the Galaxy},
  author={Dickey, John M and Lockman, Felix J},
  journal={Annual Review of Astronomy and Astrophysics},
  volume={28},
  pages={215--261},
  year={1990},
  doi={10.1146/annurev.aa.28.090190.001331}
}

@ARTICLE{wenger2024revisiting,
       author = {{Wenger}, Trey V. and {Rybarczyk}, Daniel R. and {Stanimirovi{\'c}}, Sne{\v{z}}ana},
        title = "{Revisiting the Vertical Distribution of H I Absorbing Clouds in the Solar Neighborhood}",
      journal = {Astrophysical Journal},
         year = 2024,
        month = may,
       volume = {966},
       number = {2},
          eid = {206},
        pages = {206},
          doi = {10.3847/1538-4357/ad3923},
archivePrefix = {arXiv},
       eprint = {2403.18981},
 primaryClass = {astro-ph.GA},
       adsurl = {https://ui.adsabs.harvard.edu/abs/2024ApJ...966..206W}
}

@ARTICLE{Bland-Hawthorn2016Galaxy,
       author = {{Bland-Hawthorn}, Joss and {Gerhard}, Ortwin},
        title = "{The Galaxy in Context: Structural, Kinematic, and Integrated Properties}",
      journal = {Annual Review of Astronomy and Astrophysics},
         year = 2016,
        month = sep,
       volume = {54},
        pages = {529-596},
          doi = {10.1146/annurev-astro-081915-023441},
archivePrefix = {arXiv},
       eprint = {1602.07702},
 primaryClass = {astro-ph.GA},
       adsurl = {https://ui.adsabs.harvard.edu/abs/2016ARA&A..54..529B}
}

@ARTICLE{Li2018three,
       author = {{Li}, Linlin and {Shen}, Shiyin and {Hou}, Jinliang and {Yuan}, Haibo and {Xiang}, Maosheng and {Chen}, Bingqiu and {Huang}, Yang and {Liu}, Xiaowei},
        title = "{Three-dimensional Structure of the Milky Way Dust: Modeling of LAMOST Data}",
      journal = {Astrophysical Journal},
         year = 2018,
        month = may,
       volume = {858},
       number = {2},
          eid = {75},
        pages = {75},
          doi = {10.3847/1538-4357/aabaef},
archivePrefix = {arXiv},
       eprint = {1803.10540},
 primaryClass = {astro-ph.GA},
       adsurl = {https://ui.adsabs.harvard.edu/abs/2018ApJ...858...75L}
}

@ARTICLE{diplas1994IUE,
       author = {{Diplas}, Athanassios and {Savage}, Blair D.},
        title = "{An IUE Survey of Interstellar H i Lyman-Alpha Absorption. II. Interpretations}",
      journal = {Astrophysical Journal},
         year = 1994,
        month = may,
       volume = {427},
        pages = {274},
          doi = {10.1086/174139},
       adsurl = {https://ui.adsabs.harvard.edu/abs/1994ApJ...427..274D}
}

@ARTICLE{crovisier1978kinematics,
       author = {{Crovisier}, J.},
        title = "{Kinematics of Neutral Hydrogen Clouds in the Solar Vicinity from the Nan\raisebox{-0.5ex}\textasciitildeay 21-cm Absorption Survey}",
      journal = {Astronomy and Astrophysics},
         year = 1978,
        month = nov,
       volume = {70},
        pages = {43},
       adsurl = {https://ui.adsabs.harvard.edu/abs/1978A&A....70...43C}
}

@ARTICLE{dickey2022gaskap,
       author = {{Dickey}, John M. and {Dempsey}, J.~M. and {Pingel}, N.~M. and {McClure-Griffiths}, N.~M. and {Jameson}, K. and {Dawson}, J.~R. and {D{\'e}nes}, H. and {Clark}, S.~E. and {Joncas}, G. and {Leahy}, D. and {Lee}, Min-Young and {Miville-Desch{\^e}nes}, M.-A. and {Stanimirovi{\'c}}, S. and {Tremblay}, C.~D. and {van Loon}, J. Th.},
        title = "{GASKAP Pilot Survey Science. II. ASKAP Zoom Observations of Galactic 21 cm Absorption}",
      journal = {Astrophysical Journal},
         year = 2022,
        month = feb,
       volume = {926},
       number = {2},
          eid = {186},
        pages = {186},
          doi = {10.3847/1538-4357/ac3a89},
archivePrefix = {arXiv},
       eprint = {2111.04545},
 primaryClass = {astro-ph.GA},
       adsurl = {https://ui.adsabs.harvard.edu/abs/2022ApJ...926..186D}
}

@ARTICLE{savage1996interstellar,
       author = {{Savage}, Blair D. and {Sembach}, Kenneth R.},
        title = "{Interstellar Abundances from Absorption-Line Observations with the Hubble Space Telescope}",
      journal = {Annual Review of Astronomy and Astrophysics},
         year = 1996,
        month = jan,
       volume = {34},
        pages = {279-330},
          doi = {10.1146/annurev.astro.34.1.279},
       adsurl = {https://ui.adsabs.harvard.edu/abs/1996ARA&A..34..279S}
}

@ARTICLE{werk2019nature,
       author = {{Werk}, Jessica K. and {Rubin}, K.~H.~R. and {Bish}, H.~V. and
         {Prochaska}, J.~X. and {Zheng}, Y. and
         {O{\textquoteright}Meara}, J.~M. and {Lenz}, D. and {Hummels}, C. and
         {Deason}, A.~J.},
        title = "{The Nature of Ionized Gas in the Milky Way Galactic Fountain}",
      journal = {Astrophysical Journal},
         year = 2019,
        month = dec,
       volume = {887},
       number = {1},
          eid = {89},
        pages = {89},
          doi = {10.3847/1538-4357/ab54cf},
archivePrefix = {arXiv},
       eprint = {1904.11014},
 primaryClass = {astro-ph.GA},
       adsurl = {https://ui.adsabs.harvard.edu/abs/2019ApJ...887...89W}
}

@ARTICLE{qu2020circumgalactic,
       author = {{Qu}, Zhijie and {Bregman}, Joel N. and {Hodges-Kluck}, Edmund and {Li}, Jiang-Tao and {Lindley}, Ryan},
        title = "{The Warm Gas in the MW: A Kinematical Model}",
      journal = {Astrophysical Journal},
         year = 2020,
        month = may,
       volume = {894},
       number = {2},
          eid = {142},
        pages = {142},
          doi = {10.3847/1538-4357/ab774e},
archivePrefix = {arXiv},
       eprint = {2002.06434},
 primaryClass = {astro-ph.GA},
       adsurl = {https://ui.adsabs.harvard.edu/abs/2020ApJ...894..142Q}
}

@ARTICLE{tumlinson2005hot,
       author = {{Tumlinson}, Jason and {Fang}, Taotao},
        title = "{Hot Baryons and the Distribution of Metals in the Intergalactic Medium}",
      journal = {Astrophysical Journal},
         year = 2005,
        month = apr,
       volume = {623},
       number = {2},
        pages = {L97-L100},
          doi = {10.1086/430142},
archivePrefix = {arXiv},
       eprint = {astro-ph/0501543},
 primaryClass = {astro-ph},
       adsurl = {https://ui.adsabs.harvard.edu/abs/2005ApJ...623L..97T}
}

@ARTICLE{rao2006damped,
       author = {{Rao}, Sandhya M. and {Turnshek}, David A. and {Nestor}, Daniel B.},
        title = "{Damped Ly{\ensuremath{\alpha}} Systems at z<1.65: The Expanded Sloan Digital Sky Survey Hubble Space Telescope Sample}",
      journal = {Astrophysical Journal},
         year = 2006,
        month = jan,
       volume = {636},
       number = {2},
        pages = {610-630},
          doi = {10.1086/498132},
archivePrefix = {arXiv},
       eprint = {astro-ph/0509469},
 primaryClass = {astro-ph},
       adsurl = {https://ui.adsabs.harvard.edu/abs/2006ApJ...636..610R}
}

@ARTICLE{napolitano2025composite,
       author = {{Napolitano}, Lucas and {Myers}, Adam D. and {Tedeschi}, Adam and others},
        title = "{The Composite Spectrum of QSO Absorption Line Systems in DESI DR2}",
      journal = {arXiv e-prints},
         year = 2025,
        month = dec,
          eid = {arXiv:2512.02992},
        pages = {arXiv:2512.02992},
          doi = {10.48550/arXiv.2512.02992},
archivePrefix = {arXiv},
       eprint = {2512.02992},
 primaryClass = {astro-ph.GA},
       adsurl = {https://ui.adsabs.harvard.edu/abs/2025arXiv251202992N}
}



\end{multicols}
\end{document}